\documentclass[a4paper,english,twocolumn,floatfix,amsfonts,amssymb,superscriptaddress]{revtex4-1}
\usepackage{graphicx}
\usepackage{epstopdf} 
\usepackage{dcolumn}
\usepackage{bm}
\usepackage{bbm,bbold}
\usepackage{color}
\usepackage{xcolor, soul}
\sethlcolor{green}
\usepackage{amsfonts}
\usepackage{amsmath}
\usepackage{mdframed}
\usepackage[normalem]{ulem}
\usepackage{mathrsfs}   
\usepackage[percent]{overpic}
\usepackage[colorlinks=true,citecolor=black]{hyperref}
\hypersetup{colorlinks=true,citecolor=blue,linkcolor=blue,urlcolor=blue}
\DeclareRobustCommand{\rchi}{{\mathpalette\irchi\relax}}
\newcommand{\irchi}[2]{\raisebox{\depth}{$#1\chi$}} 

\DeclareMathAlphabet\mathbfcal{OMS}{cmsy}{b}{n}

\begin{document}
\title{Light-induced shear phonon splitting and instability in bilayer graphene}
\author{Habib Rostami}
\email{hr745@bath.ac.uk} 
\affiliation{Department of Physics, University of Bath, Claverton Down, Bath BA2 7AY, United Kingdom}
\affiliation{Nordita, KTH Royal Institute of Technology and Stockholm University, Hannes Alfvéns väg 12, 10691 Stockholm, Sweden}
\date{\today}
\begin{abstract}
Coherent engineering of landscape potential in crystalline materials is a rapidly evolving research field. Ultrafast optical pulses can manipulate low-frequency shear phonons in van der Waals layered materials through the dynamical dressing of electronic structure and photoexcited carrier density. In this work, we provide a diagrammatic formalism for nonlinear Raman force and implement it to shear phonon dynamics in bilayer graphene. We predict a controllable splitting of double degenerate shear phonon modes due to light-induced phonon mixing and renormalization according to a coherent nonlinear Raman force mechanism.
Intriguingly, we obtain a light-induced shear phonon softening that facilitates structural instability at a critical field amplitude for which the shear phonon frequency vanishes. 
The phonon splitting and instability strongly depend on the laser intensity, frequency, chemical potential, and temperature of photoexcited electrons. This study motivates future experimental investigation of the optical fine-tuning and regulation of shear phonons and layer stacking order in layered van der Waals materials.

\end{abstract}
\maketitle
\section{Introduction}

 Exotic emergent phenomena in quantum systems can be generated via photoexcitation by ultrafast optical drives \cite{Ishioka_prb_2008,Murakami_prb_2015,Hohenberg_Halperin_rmp_1977,Dolgirev_prb_2020,Sun_prx_2020}. Depending on the intensity of the pump laser, we can excite and disentangle collective modes, switch the macroscopic phase of the system, dynamically engineer critical phenomena, and render robust nonlinear couplings among the different degrees of freedom in the quantum materials \cite{Huber_prl_2014,Giorgianni2019,Torre_rmp_2021}. Optical switching and photoinduced transitions correspond to the dynamical modification of the free energy landscape that is not accessible in thermal equilibrium. Photoinduced non-thermal and coherent control of correlated and topological quantum materials \cite{Vaswani_prx_2020} is being under investigation in multiple ways, such as Floquet-Bloch dressed single-particle states \cite{Oka_prb_2009} and optical dressing of many-body interaction couplings \cite{Katsnelson_prl_2015,Aoki_rmp_2014}. 
Manipulating and fine-tuning the structural phase of quantum materials by ultrashort laser pulses open a pathway to regulate quantum devices. For instance, substantial lattice deformations are reported induced by intense mid-infrared optical pulse irradiation, e.g., dynamically generated ferroelectricity and shear strain \cite{Lejman_nc_2014,Nova_science_2019}. Large photoinduced deformations are due to resonance with a vibration mode, strong Raman force, and nonlinear phonon couplings \cite{Lejman_nc_2014,Nova_science_2019,Horstmann_nature_2020,Disa_np_2021,Torre_rmp_2021,Henstridge_np_2022}.

Shear phonons in bilayer and multi-layer of 2D materials, such as the family of graphene, transition metal dichalcogenides (TMDs) and hexagonal Boron Nitride (hBN), correspond to the lateral sliding of atomic layers on each other\cite{Tan_nm_2012,Ferrari_nn_2013, Zhang_prb_2013,Zeng_prb_2012,Michel_prb_2008,Michel_prb_2012,Zhao_nl_2013,Wang_jpcc_2017, Pizzi_acsnano_2021}. Shear phonon excitation can coherently alter the staking order of layers \cite{Zhang_prx_2019,Ji_acsnano_2021}, and the electronic topology \cite{Sie_nature_2019}. Light-induced displacive dynamics \cite{born1955dynamical,lanzani2007coherent,Dekorsy_book_2000,Hase_nature_2003,Ishioka_apl_2006,Zeiger_prb_1992,Pfeifer_1992,Kuznetsov_prl_1994,Kuznetsov_prb_1995,Stevens_prb_2002,Garrett_prl_1996,Merlin_ssc_1997} of coherent shear phonons in van der Waals (vdW) layered materials such as multilayer graphene, WTe$_2$, and MoTe$_2$ \cite{Zhang_lsa_2020,Sie_nature_2019,Zhang_prx_2019,Cilento_prr_2019,Fukuda_apl_2020,Hein2020,Ji_acsnano_2021} is a promising nondestructive mechanism for controlling 2D materials properties. 
The shear mode in bilayer graphene is a double degenerate Raman-active optical mode and it has a low-frequency $\hbar\Omega_0\approx3.9$meV due to the weak vdW interlayer force \cite{Tan_nm_2012}. 
The energy and the intensity of the Raman peak for the shear mode (the C peak) strongly depend on the number of layers and inter-layer coupling. Accordingly, the spectroscopy of interlayer Raman modes is an effective method for determining layer numbers and stacking configurations, and it provides a unique opportunity to explore interlayer couplings. Driving coherent shear phonon in MoTe$_2$ causes a first-order phase transition from an inversion symmetric 1T$'$ structure to the non-centrosymmetric 1T$_d$ phase \cite{Zhang_prx_2019,Fukuda_apl_2020}. Time and angle-resolved photoemission spectroscopy (tr-ARPES) of the Weyl semimetal T$_d$-WTe$_2$, indicates coherent shear phonon-mediated control of the electronic structure \cite{Hein2020}. An optical switching from an ABA to ABC stacking is experimentally obtained by laser irradiation on trilayer graphene \cite{Zhang_lsa_2020} that might be because of the coherent shear phonon excitation. 

This paper studies the dynamical engineering of lattice potential for the shear dynamics in vdW layered materials caused by a linear polarised light field ${\bf E}(t)$. The impact of second and third-order Raman susceptibilities gives rise to light-induced corrections to the lattice potential: 
\begin{equation}
U  = \frac{1}{2} \sum_{\alpha\beta}[\Omega^2_0 \delta_{\alpha\beta}-{\cal G}_{\alpha\beta} ({\bf E})]Q_\alpha Q_\beta - \sum_\alpha {\cal F}^{(2)}_\alpha ({\bf E})  Q_\alpha,
\end{equation}
where ${\bf Q}=(Q_x,Q_y)$ is the shear phonon displacement with $\Omega_0$ being the unperturbed phonon frequency. Displacive Raman shear force described as a second-order effect $\mathbfcal{F}^{(2)}\propto EE^\ast$ in bilayer graphene has been previously investigated \cite{rostami_prb_2022}. Here, we define the third-order Raman shear force as $\mathbfcal{F}^{(3)}= \overset{\leftrightarrow}{\mathbfcal{G}}\cdot {\bf Q}\propto Q EE^\ast$ which can renormalize shear phonons and lead to a mode splitting. In particular, it can cause the instability of atomic layers to slide and form stable or metastable phases with different layer-stacking orders due to the softening of shear phonon frequency under the influence of the light field. 
The ${\cal G}_{\alpha\beta}$ coupling can be interpreted as a light-induced self-energy correction $\Sigma_{\alpha\beta}({\bf E})=-{\cal G}_{\alpha\beta}({\bf E})/2\Omega_0$ to the phonon's dynamical matrix. 
As the central result, here, we develop a diagrammatic formalism to model the impact of third-order Raman force (or light-induced phonon self-energy) on the displacive dynamics of shear phonons in layered materials. We obtain a dynamical renormalization of the shear phonons by incident light intensity leading to the splitting of the double degenerate shear phonons. We predict a lattice instability where the shear phonon frequency vanishes at a critical field amplitude. We show that the field-induced phonon splitting and instability are highly tunable by the incident laser intensity, frequency at given electronic doping, and temperature. 
Our theoretical model based on the non-equilibrium Green's function can be systematically employed in {\em ab initio} computations to study the optical engineering of shear phonon in layered materials.
 
 The rest of the paper is structured in four sections. In Section \ref {sec:method}, we provide details of the diagrammatic method for the third-order Raman force and develop a perturbative theory for optically dressed phonon's dynamical matrix. In Section \ref {sec:epc}, we summarize the mixed couplings of electrons, phonons, and photons in addition to light-matter and electron-phonon couplings in bilayer graphene. In Section \ref{sec:results}, we discuss numerical results for the light-induced phonon renormalization and, thus, its effect on the optical modulation of the shear phonon spectral function, shear mode splitting, and the light-induced shear instability. Finally, we summarize our theoretical finding, discuss it in connection with experiments, and highlight the implication of light-induced phonon renormalization in other heterostructures of 2D materials.   

\section{Method}\label{sec:method}
Stimulated Raman effect is an efficient mechanism to excite Raman-active vibrational modes \cite{dresselhaus2007group}.
The dipole moment of Raman-active phonon is linearly proportional to the light field $ \mu_b =  \alpha_{bc}  E_c$ where the polarizability tensor $\alpha_{bc}$ depends on the phonon displacement vector ${\bf Q}$. The electromagnetic potential energy thus follows $U=-\mu_bE_b= - \alpha_{bc} E_b E_c$. The corresponding Raman force driving atoms to oscillate follows a second-order nonlinear process  \cite{dresselhaus2007group}
\begin{align}
{\cal F}^{(2)}_a= -\left[\frac{\partial U}{\partial Q_a}\right]_{Q\to 0} =  
\sum_{bc} \left[\frac{\partial\alpha_{bc}}{\partial Q_a}\right]_{Q\to 0} E_b E_c.
\end{align}
Therefore the lowest-order Raman force is finite as long as the Raman susceptibility is non-vanishing, i.e., $\sigma^{(2)}_{abc}=\partial\alpha_{bc}/\partial Q_a\neq 0$. 
For large displacement, the higher-order Raman force should also be considered, which can dramatically impact phonon renormalization and lattice dynamics. The leading higher-order Raman force depends linearly on the phonon displacement and quadratically on the light field. Therefore, it is described by a third-order nonlinear mechanism
\begin{align}
{\cal F}^{(3)}_a    
 = \sum_{bcd}\left[\frac{\partial^2 \alpha_{cd}}{\partial Q_a\partial Q_b} \right]_{Q\to0} Q_b E_c E_d.
\end{align}
Formally, we have ${\cal F}^{(3)}_a = {\cal G}_{ab} Q_b$ in which ${\cal G}_{ab}$ generates a phonon self-energy in terms of a third-order Raman susceptibility $\sigma^{(3)}_{abcd}=\partial^2 \alpha_{cd}/\partial Q_a\partial Q_b$ and the incident light intensity. An anisotropic ${\cal G}_{ab}$ breaks the degeneracy of Cartesian shear modes and renormalizes the phonon's frequency and linewidth. 

To model coherent shear phonons in bilayer systems, we first provide a general theory for the Raman force and phonon self-energy using the Green's function method and diagrammatic framework. We decompose the total Hamiltonian of the system in different parts ${\cal H} = {\cal H}_e+{\cal H}_p + {\cal H}_{e-p}+{\cal H}_{lm}$ which consists of electronic kinetic Hamiltonian ${\cal H}_e$, harmonic phonon Hamiltonian ${\cal H}_p$, electron-phonon interaction
${\cal H}_{ep}$ and finally the light-matter interaction ${\cal H}_{lm}$. 
The electronic kinetic Hamiltonian reads $\hat{\cal H}_e = \sum_{\bf p} \hat \psi^\dagger_{\bf p} \hat {\cal H}({\bf p}) \hat \psi_{\bf p}$ where $\hat \psi_{\bf p}$ is the fermion annihilation spinor field at momentum $\bf p$. The harmonic shear phonon Hamiltonian with zero momentum $\bf q=0$ can be written in terms of ladder operators $\hat{\cal H}_p = \sum_{\lambda} \hbar\Omega_{0} \hat b^\dagger_\lambda \hat b_\lambda$ 
where $\hat b_\lambda $ is the phonon annihilation operator. We only consider the zone center phonon modes with a vanishing wave vector ${\bf q=0}$, and thus the phonon displacement vector is defined as 
\begin{align}
\hat Q_\lambda = \sqrt{\frac{\hbar}{\rho S \Omega_0}} (\hat b_\lambda+ 
\hat b^\dagger_\lambda)
\end{align}
in which $\lambda=x,y$ indicates two Cartesian mode components, note that $S$ stands for the area of 2D material, and $\rho$ is the mass density. Including both one-phonon and two-phonon couplings to electrons, the electron-phonon interaction Hamiltonian follows
\begin{align}
\hat{\cal H}_{e-p} &= 
\sum_{\bf p} \sum_{a} \hat\psi^\dagger_{\bf p} \hat {\cal M}^{(1)}_a ({\bf p}) \hat\psi_{\bf p} \hat Q_{a}
\nonumber\\
&+\sum_{\bf p} \sum_{ab} \hat\psi^\dagger_{\bf p} \hat {\cal M}^{(2)}_{ab} ({\bf p}) \hat\psi_{\bf p}\hat Q_{a}\hat Q_{b}.
\end{align}
Note that $\hat {\cal M}^{(1)}_a $ and $\hat {\cal M}^{(2)}_{ab}$ stand for the one- and two-phonon-electron couplings' matrix elements, respectively.   
Utilizing this effective lattice potential and the Heisenberg equation of motion, we obtain the equation of motion for coherent phonon displacement amplitude $Q_a$: 
\begin{align}
&\frac{\partial^2 Q_a(t)}{\partial t^2} +  \Gamma_p \frac{\partial Q_a(t)}{\partial t}  + \Omega^2_0 Q_a(t) 
 = \frac{{\cal F}_a(t)}{\rho}   \nonumber\\ & + \frac{1}{\rho}\sum_{b}  {\cal G}^{ins.}_{ab}(t)Q_{b}(t)
 + \frac{1}{\rho}\sum_{b}\int dt'{\cal G}^{ret.}_{ab}(t,t')Q_{b}(t') 
\end{align}
where $\Omega_0$ is the shear phonons frequency, $\Gamma_p$ stands for the 
phenomenological damping frequency of phonons.  
The leading-order Raman force is given as the expectation value of the one-phonon coupling to electrons: 
\begin{equation}
{\cal F}^{(2)}_{a}(t)  = - \frac{1}{S}\sum_{\bf p} \Big\langle \hat\psi^\dagger_{\bf p} \hat {\cal M}^{(1)}_a ({\bf p}) \hat\psi_{\bf p} \Big\rangle\Big|_{\bf Q\to0}. 
\end{equation}

The nonlinear force reveals two different dynamical forms of the light-induced phonon self-energy term ${\cal G}_{ab}$ that we label as 
{\em instantaneous} ${\cal G}^{ins.}_{ab}(t)$ and {\em retarded} ${\cal G}^{ret.}_{ab}(t,t')$ couplings.  
The instantaneous coupling is obtained as the expectation value of the two-phonon coupling matrix element 
\begin{equation}\label{eq:gins}
{\cal G}^{ins.}_{ab}(t) =- \frac{1}{S} \left[\sum_{\bf p} \left\langle \hat\psi^\dagger_{\bf p} \hat {\cal M}^{(2)}_{ab} ({\bf p}) \hat\psi_{\bf p} \right\rangle\right]_{\bf Q\to0}.
\end{equation}
While the retarded coupling is given by the variational derivative of the Raman force versus the phonon displacement field:
\begin{equation}\label{eq:gret}
{\cal G}^{ret.}_{ab}(t,t') = - \frac{1}{S}  \left[\frac{\delta}{\delta Q_{b}(t')}\sum_{\bf p}\Big\langle \hat\psi^\dagger_{\bf p} \hat {\cal M}^{(1)}_a ({\bf p}) \hat\psi_{\bf p} \Big\rangle\right ]_{\bf Q\to0}. 
\end{equation}
Note that $\langle\dots\rangle$ indicates quantum statistical averaging. 
In centrosymmetric systems, Raman-active phonons are infrared-inactive; therefore, they couple to light indirectly. The direct light-matter interaction is only through the coupling to electrons. The coupling of incident light field to electrons can be modeled by Peierls substitution ${\bf p}\to {\bf p}+e {\bf A}(t)$ in the kinetic and the electron-phonon interaction Hamiltonian terms. Considering the homogeneous vector potential ${\bf A}(t)$, the electric field reads ${\bf E}(t) = - \partial_t {\bf A}(t)$ and thus ${\bf E}(\omega) = i\omega {\bf A}(\omega)$. Formally, the light-matter interaction Hamiltonian consists of two parts: photon-electron term and photon-electron-phonon term $\hat {\cal H}_{lm}=\hat {\cal H}_{ph-e}+\hat {\cal H}_{ph-e-p}$. 
The photon-electron term follows
\begin{align}
{\cal H}_{ph-e} &= - \sum_{\bf p}  
\hat\psi^\dagger_{\bf p} 
\Big\{ \sum_a \hat j_a({\bf p}) A_a(t) 
\nonumber\\&
+ \frac{1}{2} \sum_{ab}\hat  \gamma_{ab}({\bf p})    A_a(t)  A_b(t)
+\dots\Big\}
\hat\psi_{\bf p} 
\end{align}
 where $\hat j_a$ is called the paramagnetic current operator, and $\hat \gamma_{ab}$ is known as the diamagnetic current operator as well as the Raman vertex \cite{Mahan,Devereaux_rmp_2007}. The photon-electron-phonon interaction Hamiltonian is given by the light field dependence of the electron-phonon interaction, ${\cal M}^{(1)}_a({\bf p}+e{\bf A}(t))$ and ${\cal M}^{(2)}_{ab}({\bf p}+e{\bf A}(t))$. By expanding electron-phonon matrix elements up to second-order in ${\bf A}(t)$, we obtain the photon-electron-phonon (PEP) interaction Hamiltonian $\hat {\cal H}_{ph-e-p} = 
  \sum_{\bf p} \hat\psi^\dagger_{\bf p}    \hat \Xi_{\bf p} \hat\psi_{\bf p}  $ where 
\begin{align}
& \hat \Xi_{\bf p} =  
\sum_{ab} A_a(t) Q_b(t) \Big \{ \hat \Theta^{(1)}_{ab}({\bf p})   
+\sum_{c}\hat \Theta^{(2)}_{abc}({\bf p}) Q_c(t) 
\Big \} 
\nonumber\\&+
\frac{1}{2}\sum_{abc} A_a(t) A_b(t)  Q_c(t) 
\Big 
\{ \hat \Delta^{(1)}_{abc}({\bf p}) + \sum_d \hat \Delta^{(2)}_{abcd}({\bf p})Q_d(t) \Big\}.
\end{align}

\begin{figure*}[t]
\centering
\includegraphics[width=16cm]{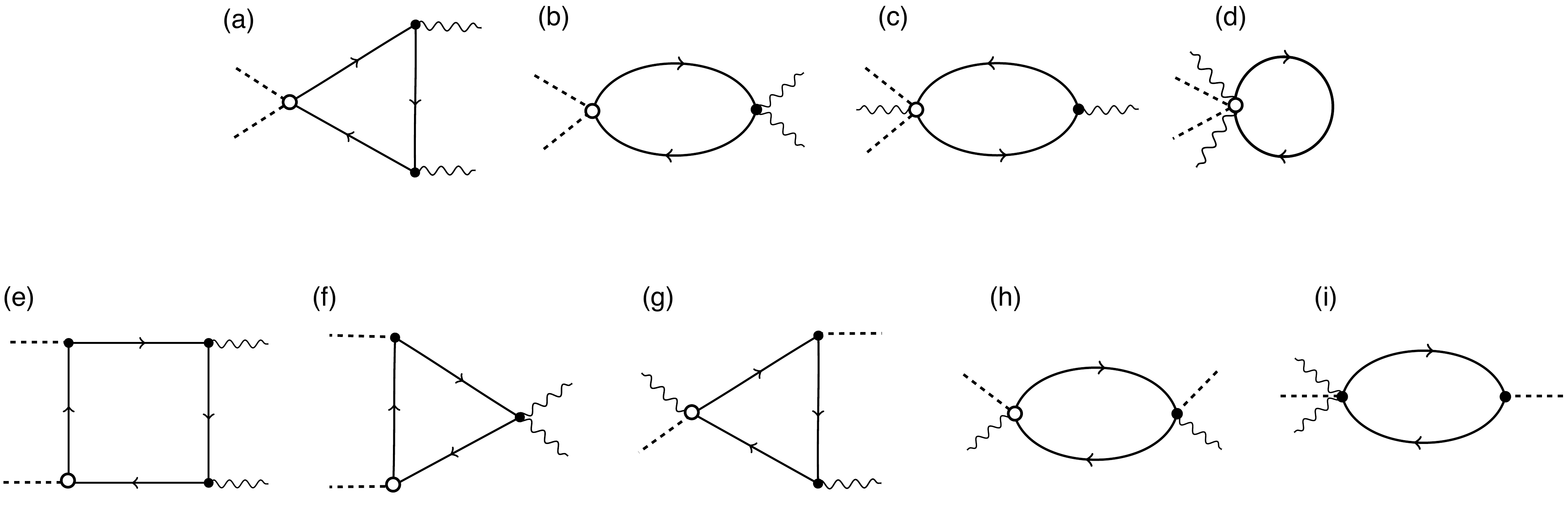}
\caption{
{\bf Feynman diagrams for light-induced phonon self-energy.} Diagrams given in panel (a-e) and (f-i) for ${\cal G}^{ins.}$ and ${\cal G}^{ret.}$ couplings, respectively. Dashed and wave lines represent external phonon and photon fields, respectively. The solid lines represent electron propagators.
}
\label{fig:diagrams}
\end{figure*}

Having defined all vertex couplings, we are equipped to evaluate the Raman force and the light-induced phonon self-energy. Because the Raman phonon is even under parity, the leading contribution to the Raman force is second order in the light field, which follows 
\begin{equation}
 {\cal F}^{(2)}_a(t) =\sum_{bc} \sum_{\omega_1,\omega_2} e^{i(\omega_1+\omega_2)t}
 \sigma^{(2)}_{abc}(\omega_1,\omega_2)  
 E_b(\omega_1) E_c(\omega_2).
\end{equation}
Similarly, the light-induced instantaneous coupling is given by  
\begin{align}
 {\cal G}^{ins.}_{ab}(t) = \sum_{cd}\sum_{\omega_1,\omega_2} e^{i(\omega_1+\omega_2)t}
 \Pi^{ins.}_{abcd}(\omega_1,\omega_2) 
 E_c(\omega_1) E_d(\omega_2).
\end{align}
Finally, one can evaluate the light-induced retarded coupling as follows 
\begin{align}
 {\cal G}^{ret.}_{ab}(t,t') &= \sum_{cd}\sum_{\omega_1,\omega_2} e^{i(\omega_1+\omega_2)t}
  \Pi^{ret.}_{abcd}(\omega_1,\omega_2,
t-t') 
\nonumber\\ &\times
 E_c(\omega_1) E_d(\omega_2).
\end{align}
Notice that $\Pi^{ret.}_{abcd}(\omega_1,\omega_2,\tau)= \sum_{\omega_3} e^{i\omega_3 \tau}\Pi^{ret.}_{abcd}(\omega_1,\omega_2,\omega_3)$ where $\omega_3$ is the phonon frequency.   
Response function $\Pi^{ins.}_{abcd}$ contributes to the instantaneous phonon self-energy since it originates from the simultaneous coupling of two phonons to electrons. On the other hand, the retarded response function $\Pi^{ret.}_{abcd}$ contains memory effects where the past dynamics of phonons can influence their future motion.

The light-induced rigid displacement directly depends on the displacive Raman force that is the rectified part of the force in a second-order nonlinear process \cite{rostami_prb_2022}. 
We consider monochromatic incident light-field ${\bf E}(t)= E_0 \hat {\bm\varepsilon} e^{-i\omega t} + c.c.$ with $\hat {\bm\varepsilon}$ being the linear polarization unit vector. The displacive force is thus given by the rectification process (i.e., $\omega_1=-\omega_2=\omega$) that leads to the following time-independent Raman force: 
\begin{align}
 {\cal F}^{\rm D}_a  &= \sum_{bc}
 \sigma^{(2)}_{abc}(\omega,-\omega)  
E_b(\omega) E^\ast_c(\omega).
\end{align}
Similarly, the rectified component of the instantaneous phonon-phonon coupling reads 
\begin{equation}
 {\cal G}^{ins.}_{ab}   =  \sum_{cd}
\Pi^{ins.}_{abcd}(\omega,-\omega)   
E_c(\omega) E^\ast_d(\omega), 
\end{equation}
and the retarded light-induced phonon-phonon coupling follows 
\begin{equation}
 {\cal G}^{ret.}_{ab}(t-t') = \sum_{cd}
 \Pi^{ret.}_{abcd}(\omega,-\omega,t-t') 
E_c(\omega) E^\ast_d(\omega).
\end{equation}
It is worth highlighting that the second harmonic parts of ${\cal G}^{ins.}_{ab}$ and ${\cal G}^{ret.}_{ab}$ $\sim e^{i2\omega t}$ do not contribute noticeably due to its convolution with the slow oscillation of the ion displacement $Q_a\sim e^{i\Omega_0 t}$ since $\omega\gg\Omega_0$. In this regard, the rectified parts of ${\cal G}^{ins.}_{ab}$ and ${\cal G}^{ret.}_{ab}$ play the dominant role. 
Eventually, the phonon equation of motion coherently dressed by the light field is given by 
\begin{align}\label{EOM}
&\frac{\partial^2 Q_a(t)}{\partial t^2}+\Gamma_p \frac{\partial Q_a(t)}{\partial t}  + \Omega^2_0 Q_a(t) 
 =  \frac{{\cal F}^{\rm D}_a}{\rho}   
 +  \frac{1}{\rho}\sum_{b} 
 {\cal G}^{ins.}_{ab} Q_{b}(t)  
 \nonumber\\ &+  \frac{1}{\rho} \sum_{b} \int dt'
 {\cal G}^{ret.}_{ab}(t-t') Q_{b}(t'). 
\end{align}
We employ a diagrammatic formalism to estimate numerical values of the Raman force \cite{rostami_prb_2022} and phonon self-energy. Here, the main focus is on the light-induced renormalization and mixing of shear phonons. The Feynman diagrams for the instantaneous and retarded couplings thus are given in Fig.~\ref{fig:diagrams}a-d and Fig.~\ref{fig:diagrams}e-i, respectively. To quantitatively analyze the spectral function and the splitting of shear phonons, we microscopically explore the coherent dynamics of shear modes in bilayer graphene in the remaining part of the paper.

\section{Light-matter and electron-phonon couplings}\label{sec:epc}
Bilayer graphene consists of two single layers
of graphene sheets offset from each other in the $xy$ plane. The low-energy quasiparticles follow a two-band Hamiltonian around the corners of the hexagonal Brillouin zone \cite{McCann_prl_2006}
\begin{align}\label{eq:Ham-2band}
\hat H_{\bf p} = -\frac{1}{2m}\{(p^2_x-p^2_y)\hat\sigma_x + 2 \tau p_x p_y \hat \sigma_y \} -\mu\hat I. 
\end{align}
Note that  ${\bm p} = \hbar {\bm k}$ is the momentum vector, $\tau=\pm$ indicates 
 two K and K$'$ valley points, the identity matrix $\hat I$ and
Pauli matrices $\hat\sigma_{x}$ and $\hat\sigma_{y}$ are in the layer pseudospin basis, and
$\mu$ is the chemical potential. In our convention, the $x$-direction shows a zigzag orientation of the honeycomb lattice \cite{Rostami_prb_2013b}. The effective mass is given by $1/2m \approx v^2 /|\gamma_1| $ with $v\approx 10^6$m/s and vertical inter-layer hopping energy $\gamma_1\approx-0.4$eV \cite{McCann_rpp_2013}.  
Having the in-plane displacement ${\bf Q}^{(\ell)}({\bf r})$ of two layers $\ell=1,2$, the shear phonon displacement is the asymmetric component: 
\begin{align}
{\bf Q}= \frac{{\bf Q}^{(1)}-{\bf Q}^{(2)}}{\sqrt{2}}.
\end{align}
The shear displacement vector is even under parity ${\cal P}$ since ${\cal P} \{{\bf Q}^{(1)},{\bf Q}^{(2)}\}{\cal P}^{-1} = -\{{\bf Q}^{(2)},{\bf Q}^{(1)}\}$ leading to ${\cal P} {\bf Q}{\cal P}^{-1}= {\bf Q}$. Therefore, the shear mode is a Raman-active but IR-inactive phonon. 
We consider the coupling of electrons to one and two photons given by $ \hat j_\alpha = -e \partial_{p_\alpha} \hat H_{\bf p} $ and $\hat \gamma_{\alpha\beta} = - e^2 \partial_{p_\alpha}\partial_{p_\beta} \hat H_{\bf p}$, respectively. The coupling of electrons to one and two photons are thus given by 
\begin{align}
&(\hat j_x,\hat j_y) 
= \frac{e}{m} (p_x\hat\sigma_x+\tau p_y\hat\sigma_y,-p_y\hat\sigma_x+\tau p_x\sigma_y),
\nonumber\\
&(\hat \gamma_{xx}=-\hat \gamma_{yy},\hat \gamma_{xy}=\hat \gamma_{yx}) =  \frac{e^2}{m}(\hat\sigma_x,\tau \hat\sigma_y).
\end{align}

The electron-phonon couplings are obtained using a four-band tight-binding model following the approach developed in Ref. \cite{rostami_prb_2022} providing the detailed analysis of electron coupling to shear phonons in bilayer graphene using tight-binding and ${\bf k\cdot p}$ models-- see also Refs. \cite{Ishikawa_2006, Cappelluti_prb_2012,Basko_2009}. Accordingly, the couplings of electrons to shear phonons in the low-energy model read  \cite{rostami_prb_2022}
\begin{align}\label{eq:electron-phonon}
&(\hat {\cal M}^{(1)}_x,\hat {\cal M}^{(1)}_y)\approx {\cal M}^{(1)} (\tau\hat \sigma_y, \hat\sigma_x),
\nonumber\\  
&(\hat {\cal M}^{(2)}_{xx}=-\hat {\cal M}^{(2)}_{yy},\hat {\cal M}^{(2)}_{xy}=\hat {\cal M}^{(2)}_{yx})  \approx {\cal M}^{(2)} (\hat\sigma_x,\tau \hat\sigma_y). 
\end{align}
Electron-phonon coupling can depend on the light field, and this leads to mixed PEP couplings which are obtained after neglecting electron momentum $p$ \cite{rostami_prb_2022}
\begin{align}
&
\begin{pmatrix}
\hat \Theta^{(1)}_{xy}= \hat\Theta^{(1)}_{yx} \\ 
\hat\Theta^{(1)}_{yy}=- \hat\Theta^{(1)}_{xx} 
\end{pmatrix} \approx  -\Theta^{(1)}
\begin{pmatrix}
\tau \hat\sigma_x \\  \hat\sigma_y
\end{pmatrix}, 
\\&
\begin{pmatrix}
 \Theta^{(2)}_{yyx}=\Theta^{(2)}_{yxy}=
\Theta^{(2)}_{xyy}=\Theta^{(2)}_{xxx}/3
 \\
\Theta^{(2)}_{xyx}=\Theta^{(2)}_{xxy}=
\Theta^{(2)}_{yxx}=\Theta^{(2)}_{yyy}/3 
\end{pmatrix}
\approx\Theta^{(2)}
\begin{pmatrix}
 \tau\hat\sigma_x
\\
-\hat\sigma_y
\end{pmatrix}, 
\\&
\begin{pmatrix}
\Delta^{(1)}_{xxy}=\Delta^{(1)}_{xyx}=\Delta^{(1)}_{yxx}=\Delta^{(1)}_{yyy}/3
 \\
\Delta^{(1)}_{yxy}=\Delta^{(1)}_{yyx}=\Delta^{(1)}_{xyy}=\Delta^{(1)}_{xxx}/3
\end{pmatrix}
\approx \Delta^{(1)}
\begin{pmatrix}
\hat\sigma_x
\\
\tau\hat\sigma_y
\end{pmatrix}. 
\end{align}
The expression for $\Delta^{(2)}_{abcd}$ coupling, representing the coupling of two-photon and two-phonon fields with an electron field, has yet to be specified. However, we can include its contribution using a gauge invariance argument, and therefore there is no need to explicitly calculate $\Delta^{(2)}_{abcd}$ coupling constants. This gauge invariance issue is discussed more explicitly in the following sections.

The values of electron-phonon couplings strength are given in terms of microscopic parameters of the system \cite{rostami_prb_2022}
\begin{align}
&{\cal M}^{(1)} = -\Big(\frac{3 a_0}{\sqrt{2} b}\Big)\Big( \frac{\partial\gamma_3}{\partial b}\Big)
=  \Big(\frac{3 a_0 \gamma_3}{\sqrt{2} b^2}\Big) \beta_3,
\\
&{\cal M}^{(2)} = \Big(\frac{3 a^2_0}{4b^2} \Big) 
\Big(\frac{\partial^2\gamma_3}{\partial b^2} \Big)
=
 \Big(\frac{3 a^2_0 \gamma_3}{4b^4} \Big) \beta_3(1+\beta_3).
\end{align}
where $\gamma_3 \approx 0.3$eV \cite{McCann_rpp_2013} is an interlayer hopping energy corresponding to the hopping of electrons from sublattice A of bottom layer one to sublattice A of the top layer in a Bernal stack bilayer system. The Gruneisen parameter follows $\beta_3 = - \partial \ln \gamma_3/\partial \ln b$. The mixed PEP coupling constants are thus obtained as
$\Theta^{(1)}= 
 ({ea_0}/{2\hbar}){\cal M}^{(1)}$,  $\Delta^{(1)}= 
 ({ea_0}/{2\hbar})^2{\cal M}^{(1)}$ and $\Theta^{(2)} = - 
({e a_0}/{4\hbar}) {\cal M}^{(2)}$.
Accordingly, the only coupling parameter is $\beta_3$. 
In the second equality of the above relation, we assume the power-law rule $\gamma_3 \sim 1/b^{\beta_3}$ for the dependence of $\gamma_3$ on the corresponding bond length $b=\sqrt{a^2_0+c^2}\approx 0.38$nm with intralayer Carbon-Carbon bond length $a_0=0.142$nm and the interlayer distance $c=0.34$nm. 
An analysis based on the density functional calculation estimates the dependence of $\gamma_3$ on the bond length as $\partial \gamma_3/\partial b\approx -0.54 {\rm eV/\AA}$ \cite{Cappelluti_prb_2012} and therefore one can obtain the Gruneisen parameter $\beta_3=-(b/\gamma_3)\partial \gamma_3/\partial b\approx 6.84$. The vertical hopping derivative ${\partial\gamma_1}/{\partial c}$ does not contribute to the leading order electron-phonon interaction. 
 
\section{Numerical results and discussion} \label{sec:results}
The phonon self-energy terms depend on $\Pi^{ins.}_{abcd}$ and $\Pi^{ret.}_{abcd}$ which are given in terms of corresponding susceptibilities $\rchi^{ins.}_{abcd}$ and $\rchi^{ret.}_{abcd}$ in response to the vector potentials $A_c(\omega_1)$ and $A_d(\omega_2)$. Therefore, we have 
\begin{align}
&\Pi^{ins.}_{abcd}(\omega_1,\omega_2) = - \frac{\rchi^{ins.}_{abcd}(\omega_1,\omega_2)}{(i\omega_1)(i\omega_2)}, 
\\
&\Pi^{ret.}_{abcd}(\omega_1,\omega_2,\omega_3) = - \frac{\rchi^{ret.}_{abcd}(\omega_1,\omega_2,\omega_3)}{(i\omega_1) (i\omega_2)}. 
\end{align}
where $\omega_3$ is the phonon frequency. 
The overall minus sign in the above relations by definition, given in Eq.~(\ref{eq:gins}) and Eq.~(\ref {eq:gret}). Utilizing Feynman diagrams given in Fig.~\ref{fig:diagrams}(a-d) and Fig.~\ref{fig:diagrams}(e-i) we calculate $\rchi^{ins.}_{abcd}$ and $\rchi^{ret.}_{abcd}$ response functions, respectively. Before reporting the numerical results, it is necessary to mention that the contribution of the mixed PEP coupling $\Delta^{(2)}_{abcd}$, depicted in the diagram in Fig.~\ref{fig:diagrams}d, is frequency-independent. One can incorporate this diagram by enforcing the gauge invariance, implying a vanishing system response to a static homogeneous gauge field. Accordingly, we have $\rchi^{ins.}_{abcd}(\omega_1=0,\omega_2=0)+\rchi^{ret.}_{abcd}(\omega_1=0,\omega_2=0,\omega_3)=0$ so that the impact of $\Delta^{(2)}_{abcd}$ can be taken into account by subtracting the static value of each diagram. 
\begin{align}
&\rchi^{ret.}_{abcd} (\omega_1,\omega_2,\omega_3) \to \rchi^{ret.}_{abcd} (\omega_1,\omega_2,\omega_3)-\rchi^{ret.}_{abcd} (0,0,\omega_3),
\nonumber\\ 
&\rchi^{ins.}_{abcd} (\omega_1,\omega_2)\to \rchi^{ins.}_{abcd} (\omega_1,\omega_2)-\rchi^{ins.}_{abcd} (0,0).
\end{align}
The rest of the section summarises our analytical and numerical results for the light-induced instantaneous and retarded couplings and the resulting renormalization of shear phonon frequency in bilayer graphene. Afterward, we quantitatively analyze the splitting of shear phonon modes and phonon instability which are coherently controllable by altering the incident laser intensity, frequency, and polarization. We also investigate the impact of finite electronic temperature on our numerical results. 
\begin{figure*}[t]
\centering
\includegraphics[width=18cm]{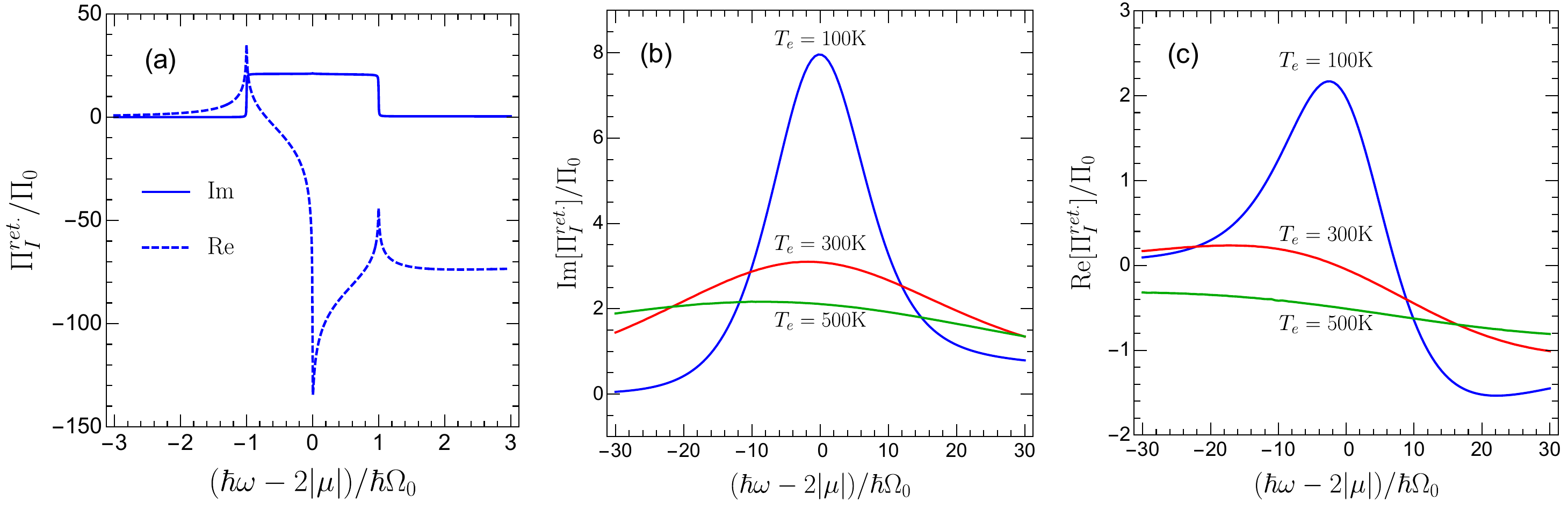}
\caption{{\bf Light-induced retarded self-energy coupling versus the laser frequency.}  Panel (a) indicates the imaginary and real parts of $\Pi^{ret.}_I$ versus the incident laser frequency at zero electronic temperature $T_e=0$ and $\Gamma_e=0.001$meV. Panel (b) and (c) respectively illustrate the imaginary and real parts of $\Pi^{ret.}_I$ at $\Gamma_e=1$meV and different values of electronic temperature $T_e$. We set $\mu=200$meV, and $\hbar\Omega_0=3.9$meV in this figure.} 
\label{intermode_coupling}
\end{figure*}
\subsection{Light-induced instantaneous self-energy}
Light-induced instantaneous phonon self-energy is calculated following the Feynman diagrams depicted in Fig.~\ref{fig:diagrams}a-d employing the effective low-energy description of electrons and the couplings to phonons and photons. For a linearly polarized incident light field ${\bf E}(t) = E_0 (\hat{\bf x}\cos\theta+\hat{\bf y}\sin\theta) e^{-i\omega t}+c.c$, the symmetry of the system enforces the following constraints for the only non-vanishing tensor elements as $-\Pi^{ins.}_{xxxx}=-\Pi^{ins.}_{yyyy}= \Pi^{ins.}_{xxyy}=\Pi^{ins.}_{yyxx}= \Pi^{ins.}_{xyxy}=\Pi^{ins.}_{yxyx}= \Pi^{ins.}_{xyyx}=\Pi^{ins.}_{yxxy}=\Pi^{ins.}$. In accordance with this symmetry constrain, we find the dependence of $\hat {\cal G}^{ins.}$ on the light field polarization angle $\theta$: 
\begin{align}
    \hat {\cal G}^{ins.} 
    =  \Pi^{ins.}(\bar\omega_1,\bar\omega_2)    E^2_0 
    \begin{bmatrix}
    -\cos(2\theta) &\sin(2\theta) \\[5pt] \sin(2\theta) & \cos(2\theta) 
    \end{bmatrix}
\end{align}
where $\bar\omega_j = \hbar(\omega_j+i\Gamma_e)/|\mu|$ with $\omega_1=-\omega_2=\omega$ and $\hbar\Gamma_e$ is the electron scattering rate. The functional dependence of the instantaneous coupling is obtained analytically using the effective low-energy Hamiltonian, and it reads $\Pi^{ins.}(\bar\omega_1,\bar\omega_2)=  {\Pi}^{ins.}_0 \Lambda(\bar\omega_1,\bar\omega_2)$ with (see Appendix \ref{app:inst})
\begin{align}
\Lambda(\bar\omega_1,\bar\omega_2)
&= \Bigg\{ 
 \frac{ (\bar\omega_1+2\bar\omega_2)}{\bar\omega^2_2(\bar\omega_1+\bar\omega_2)} \ln\left[\frac{4-\bar\omega^2_1}{4-(\omega_1+\omega_2)^2}\right] 
\nonumber\\& 
+  \frac{ (\bar\omega_2+2\bar\omega_1)}{\bar\omega^2_1(\bar\omega_1+\bar\omega_2)} \ln\left[\frac{4-\bar\omega^2_2}{4-(\bar\omega_1+\bar\omega_2)^2}\right] 
\nonumber\\&
-\frac{3}{2\bar\omega_1\bar \omega_2}  \ln \left[1-\frac{(\omega_1+\omega_2)^2}{4}\right]
\Bigg\},
\end{align}
and the dimensionful constant prefactor reads  
\begin{align}
 {\Pi}^{ins.}_0= \frac{N_f e^2{\cal M}^{(2)} }{4\pi\mu^2}.
 \end{align}
in which $N_f=4$ stands for the spin-valley degeneracy. The properties of universal function $\Lambda(\omega_1,\omega_2)$ are explored in Ref. \cite{rostami_prb_2022} at zero and finite electronic temperature $T_e$.  

\subsection{Light-induced retarded self-energy}
For the retarded coupling, we have the Feynman diagrams depicted in Fig.~\ref{fig:diagrams}(e-i), among which only diagrams shown in panels (e), (h), and (i) of Fig.~\ref{fig:diagrams} are non-zero in our effective model analysis. The symmetry of our low-energy model results in constraints for the non-vanishing elements of $\Pi^{ret.}_{abcd}$ as $\Pi^{ret.}_{xxxx}=\Pi^{ret.}_{yyyy}$, $\Pi^{ret.}_{xxyy}=\Pi^{ret.}_{yyxx}$, $\Pi^{ret.}_{xyxy}=\Pi^{ret.}_{yxyx}$ and $\Pi^{ret.}_{xyyx}=\Pi^{ret.}_{yxxy}$. Accordingly, the polarization dependence of the retarded coupling reads 
\begin{align}
\hat {\cal G}^{ret.} (\Omega) &= E^2_0 \Pi^{ret.} _{I} \hat I
+ E^2_0 \begin{bmatrix}
\Pi^{ret.} _{Z}\cos(2\theta)  
&
\Pi^{ret.} _{X}  \sin(2\theta)
\\[5pt] 
\Pi^{ret.} _{X}  \sin(2\theta)
 &
 - \Pi^{ret.} _{Z}\cos(2\theta) 
 \end{bmatrix}, 
\end{align}
where for a given light field frequency $\omega$ and at phonon frequency $\Omega$, we define 
\begin{align}
\Pi^{ret.} _{I,Z} (\omega,-\omega,\Omega) &= \frac{\Pi^{ret.} _{xxxx} \pm \Pi^{ret.} _{xxyy} }{2},
\nonumber\\
\Pi^{ret.} _{X} (\omega,-\omega,\Omega)&= \frac{\Pi^{ret.} _{xyyx} +\Pi^{ret.} _{xyxy} }{2}. 
\end{align}
The $+$ and $-$ signs in the above relation refer to $\Pi^{ret.} _{I}$ and $\Pi^{ret.} _{Z}$, respectively. %
The three contributions from three diagrams Fig.~\ref{intermode_coupling}a,d,e can be collected as follows 
\begin{align}
&\Pi^{ret.}_{\xi=I,Z,X}(\omega_1,\omega_2,\omega_3) = \Pi^{ret.}_0 
\Big\{ \Pi^{\rm square}_{\xi} (\omega_1,\omega_2,\omega_3) +
 \nonumber\\& \alpha \left[ \Pi^{\rm bubble-\Theta}_{\xi} (\omega_1,\omega_2,\omega_3)  + \Pi^{\rm bubble-\Delta}_{\xi}(\omega_1,\omega_2,\omega_3) \right] \Big\}. 
\end{align}
In the low-energy model, we obtain vanishing contributions for the triangle diagrams shown in Fig. \ref{fig:diagrams}f,g.    
The detailed derivation and analytical expressions of the above nonlinear response functions at zero electronic temperature are given in Appendix \ref{app:ret}. Notice the constant factors  
\begin{align}
\Pi^{ret.}_0&= \frac{N_f(e{\cal M}^{(1)})^2}{24\pi (\hbar\Gamma_e) \mu^2 }
~~, ~~ 
\alpha= \frac{\hbar\Gamma_e }{(18\gamma^2_0/\gamma_1)}.
\end{align}
Since $18\gamma^2_0/\gamma_1\approx 10^2 $eV and $\hbar\Gamma_e$ is usually less than tens of meV, we have $\alpha\ll1$ for realistic value of scattering rate $\hbar\Gamma_e$. Therefore, we safely neglect the contribution of bubble diagrams relative to the square diagram. Considering the square diagram, our microscopic calculation gives $\Pi^{ret.}_{xxxx}=\Pi^{ret.}_{xxyy}$ and $\Pi^{ret.}_{xyxy}=-\Pi^{ret.}_{xyyx}$. Consequently, we obtain  
\begin{align}
\Pi^{ret.}_I=\Pi^{ret.}_{xxxx}\approx \Pi^{\rm square}_{I}~~~\text{and}~~~
\Pi^{ret.}_{Z} =\Pi^{ret.}_{X}=0.   
\end{align}

\begin{figure*}[t]
\centering
\includegraphics[width=18cm]{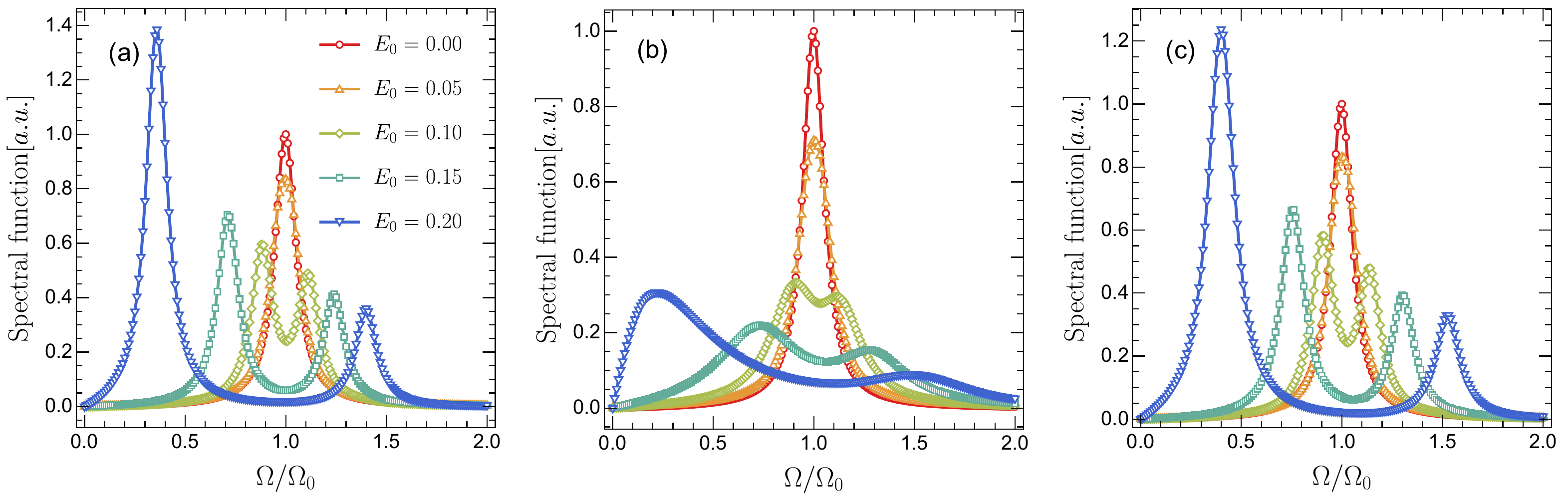}
\caption{{\bf Adiabatic and non-adiabatic spectral functions of optically dressed shear phonons.} Panels (a) and (b) illustrate the results obtained within the adiabatic and non-adiabatic models, respectively. In panel (c), we depict the spectral function obtained after neglecting the imaginary part of $\Pi^{ret}_I (\Omega)$. 
The shear phonons' splitting is depicted at different values of the light field amplitude in the unit of V/nm. The splitting is almost the same in both adiabatic and non-adiabatic models. However, the linewidth and peak values are different in the two models.   
 We set $\mu=200$meV, $\hbar\Gamma_e=5$meV, $\hbar\Gamma_p=0.5$meV, $T_e=100$K, $\hbar\Omega_0=3.9$meV and $\theta=0$ in this figure.}
\label{spectral}
\end{figure*}

In Fig. \ref{intermode_coupling}, we illustrate real and imaginary parts of $\Pi^{ret.} _{I}$ at zero and finite electronic temperature $T_e$. At very low temperatures, the imaginary part is finite only in a narrow frequency window close to the interband optical transition gap $2|\mu|$ where the width of the frequency window is given by the shear phonon frequency $2\Omega_0$.
The real part of $\Pi^{ret.}_{I}$ shows logarithmic cusps at optical transition edges for $\hbar\omega=2\mu$ and $\hbar\omega= 2\mu\pm\hbar\Omega$. 

We generalize the zero temperature response function $\Pi_{abcd}(\varepsilon_F, T_e=0,\dots)$ to finite electronic temperature using the Maldague's  formula \cite{giuliani2005quantum, rostami_prb_2022}, by integrating over the Fermi energy as follows  
\begin{align}
   \Pi_{abcd}|_{\mu,T_e}= \int^{\infty}_{-\infty} dy \frac{\Pi_{abcd}|_{\varepsilon_F\to y,T_e=0}}{4k_{B}T_e \cosh^2\left({\frac{y-\mu}{2k_B T_e}}\right)}.
\end{align}
The electronic temperature can reach thousands of Kelvin due to intense and ultrashort laser pulses \cite{Lui_prl_2010,Tomadin_prb_2013,Brida_nc_2013,Tomadin_sadv_2018,Andreatta_prb_2019}. The imaginary part of $\Pi^{ret.} _{I}$ is always positive at zero and finite temperatures. We investigate the impact of the electronic temperature, and the result shows an expected reduction of the response for frequencies in the range $|\hbar\omega-2|\mu||<\hbar\Omega_0$ while outside this range, the response function increases by raising the temperature. 

\begin{figure*}[t]
\centering
\includegraphics[width=18cm]{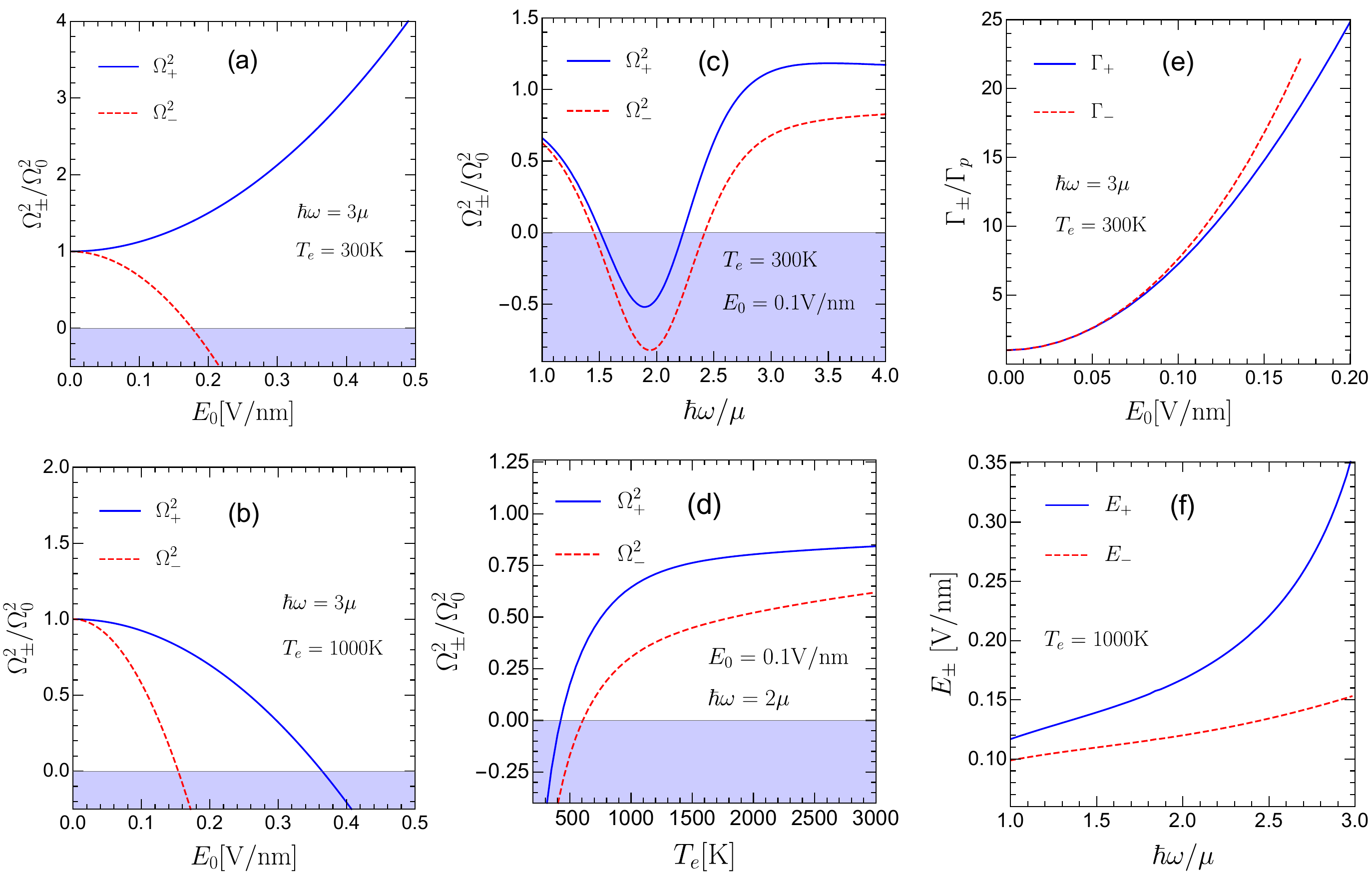}
\caption{{\bf Light-induced shear phonon splitting and instability.} 
Panel (a) and (b) shows the mode splitting as a function of the light field amplitude at laser frequency $\hbar\omega=3\mu$ and two different values of electronic temperature.
Panel (c) illustrates the frequency dependence of dynamically renormalized shear modes at field amplitude $E_0=0.1$V/nm and $T_e=300$K. 
Panel (d) indicates the shear phonon frequencies versus the electronic temperature at $\hbar\omega=2\mu$ and $E_0=0.1$V/nm. 
Panel (e) shows the field dependence of shear phonon linewidth at $\hbar\omega=3\mu$ and $T_e=300$K. 
Panel (f) manifests the laser frequency dependence of the critical field amplitude at which phonon modes become unstable. We set $\mu=200$meV, $\hbar\Gamma_p=0.1$meV, $\hbar\Gamma_e=5$meV and $\hbar\Omega_0=3.9$meV in this figure.
}
\label{splitting}
\end{figure*}

\subsection{Light-induced phonon splitting and instability}  
Performing the Fourier transform of the shear phonon displacement vector $Q_a(t) = \sum_\Omega Q_a(\Omega) e^{-i\Omega t}$ in Eq. (\ref{EOM}) leads to the equation of motion into the frequency domain  
\begin{align}
\sum_{b} \Big\{{\cal K}_{ab}(\Omega)  -(\Omega^2 + i\Gamma_p \Omega)\delta_{ab}\Big\} Q_{b}(\Omega) 
 =  \frac{{\cal F}^{\rm D}_a  }{\rho}
\end{align}
where the dynamical matrix of shear modes is dressed by the external light field and given by
\begin{align}
{\cal K}_{ab} (\Omega) = \Omega^2_0 \delta_{ab}- \frac{{\cal G}^{ins.}_{ab}}{\rho}- \frac{{\cal G}^{ret.}_{ab}(\Omega)}{\rho}. 
\end{align}
We write the light-induced phonon self-energy term in a compact form in the unit of a characteristic frequency $\nu_0 = \sqrt{g_0/\rho}$ with $g_0 = \gamma_3(eE_0/b\mu)^2$ and thus the dynamical matrix reads 
\begin{align}
\hat {\cal K}(\Omega)   = \Omega^2_0 \hat I-
 \nu^2_0 \begin{bmatrix}
K _{I}  + K _{Z} \cos(2\theta)  
&
K _{X}   \sin(2\theta)
\\[5pt]
K_{X} \sin(2\theta)
 &
K_{I}  - K_{Z} \cos(2\theta) 
 \end{bmatrix}. 
\end{align}
where for given driving field frequency $\omega$, we have
\begin{align}
&K_I(\Omega)  = r_1 \Pi^{\rm square}_I (\omega,-\omega,\Omega),
 \nonumber\\
&K_X  =-K_Z = r_0 \Lambda (\omega,-\omega).
\end{align}
Notice that $r_0 = 3 a^2_0 \beta_3(1+\beta_3)/(4\pi b^2)$ and $r_1=3 a^2_0 \beta^2_3 \gamma_3/(4\pi b^2 \hbar\Gamma_e)$ are dimensionless parameters. Considering numerical values of $\gamma_3$ and $\beta_3$ and lattice parameters, we obtain $r_0\approx 1.738$ and $r_1\approx 455/(\hbar\Gamma_e [{\rm meV}])$. Utilizing the dynamical matrix, we introduce the phonon Green's function dressed by the light field 
\begin{equation}
\hat D(\Omega)  =  \big [(\Omega^2+ i\Gamma_p \Omega)\hat I-\hat {\cal K} (\Omega) \big ]^{-1}.
\end{equation}
Therefore, the spectral function of the shear mode is defined as 
${\cal A}(\Omega)= -{\rm Im}[{\rm Tr}[\hat D(\Omega)]]/\pi$. 
By defining $\nu^2_0 \tilde K_I(\Omega) = \nu^2_0 K_I(\Omega) + i\Gamma_p \Omega$ and considering $K_X=-K_Z$, we obtain a $\theta$-independent spectral function  
\begin{equation}
{\cal A}(\Omega)= \frac{2}{\pi} {\rm Im}\left[\frac{\Omega^2_0-\Omega^2-\nu^2_0 \tilde K_I(\Omega)}
{[\Omega^2_0-\Omega^2-\nu^2_0 \tilde K_I(\Omega)]^2+\nu^4_0K^2_Z}\right]. 
\end{equation}
For displacive Raman force analysis and rigid shear displacement, we only need to know the dynamical matrix at $\Omega=0$, which corresponds to the adiabatic component of the spectral function. In the adiabatic approximation \cite{Giustino_rmp_2017}, Green's function is obtained by setting the phonon frequency to zero in the dynamical matrix ${\cal K}(\Omega=0)$: 
\begin{equation}
\hat D^{\rm ad}(\Omega) = \big [(\Omega^2+ i\Gamma_p \Omega)\hat I-\hat {\cal K} (\Omega=0) \big ]^{-1}. 
\end{equation}
We calculate the spectral function for both adiabatic and non-adiabatic models, and the results are depicted in Fig.~\ref{spectral}a,b. Both models predict a splitting of degenerate shear modes due to the impact light field based on the nonlinear Raman mechanism. This comparison shows that the adiabatic approximation nicely predicts the same value for splitting phonon modes in the non-adiabatic formalism. However, the two methods differ for the linewidth and the spectral weight peak value. In particular, in Fig.~\ref{spectral}c, we neglect the imaginary part of $\Pi^{ret}_I(\Omega)$, which results in sharper peaks coinciding with the spectral peaks in the adiabatic model. According to this analysis, we can safely consider an adiabatic approximation by setting $\Omega=0$ in the dynamical matrix $\hat{\cal K}(\Omega=0)$ to discuss the light-induced shear mode splitting and instability at which phonon frequency vanishes. In this case, we diagonalize the adiabatic dynamical matrix and obtain the normal shear phonon modes in a linear superposition of two Cartesian modes. Eventually, the normal mode frequencies read  
 \begin{equation}\label{eq:Om_pm}
 \left(\frac{\Omega_\pm}{\Omega_0}\right)^2 = 1 -\xi^2 K_I(0) 
 \pm \xi^2 |K_Z|.
 \end{equation}
Note that $\xi=\nu_0/\Omega_0$ is a dimensionless parameter, and both $K_I(0)$ and $K_Z$ are real numbers. Since $K_I(\Omega)$ is a complex number, its imaginary part induces a field-dependent renormalization of the phonon linewidth that follows 
 \begin{align}
     \frac{\Gamma_{\pm}}{\Gamma_p} =1 + \frac{\nu^2_0}{\Gamma_p \Omega_{\pm}} {\rm Im}[K_I(\Omega_\pm)]. 
 \end{align}
There are some qualitative features in the field-dependent phonon frequency and linewidth:
($i$) First, our perturbative analysis is primarily valid for small enough $\xi$; therefore, we have $\Omega_\pm>0$ in the best validity range of our formalism. However, we can predict the case $\Omega_\pm=0$ at critical field amplitudes $E_\pm $ for which the shear mode becomes unstable that can facilitate an optically driven structural phase transition of the vdW material via the change in the staking order.  
($ii$) For $K_I(0)=0$ the splitting of two modes is symmetric and $\Omega_+$ is always non-zero while $\Omega_{-}$ vanishes at $\nu_0=\Omega_0/\sqrt{|K_Z|}$. ($iii$) For $|K_Z| \ll |K_I(0)|$, phonons remain degenerate at a larger or smaller frequency relative to $\Omega_0$ for $K_I(0)<0$ or $K_I(0)>0$, respectively. If $K_I(0)>0$, phonon modes get soften ($\Omega_\pm\to0$) at  a critical field amplitude leading to $\nu_0=\Omega_0/\sqrt{K_I(0)}$. ($iv$) Since ${\rm Im}[K_I(\Omega_\pm)]>0$ as shown in Fig. \ref{spectral}c, we obtain a field-induced broadening of spectral function due to the optically enhanced electron-phonon scattering. 

We illustrate the normal mode frequency in Fig. \ref{splitting}a,b as a function of the incident field intensity at two different electronic temperature values manifesting the quadratic dependence on the field amplitude. Fig. \ref{splitting}a shows that two normal modes conversely evolve where  $\Omega_{+}$ ($\Omega_{-}$) increases (decreases) by raising the field amplitude $E_0$. 
The diverging evolution of two phonon modes' frequency in opposite directions becomes a converging trend with negative renormalization and phonon softening at higher electronic temperatures. This is because $K_I(0)$ enhances by raising the temperature, and thus, it becomes larger than $K_Z$ leading to a converging trend for both $\Omega_\pm$ versus field amplitude $E_0$. Intriguingly, at a critical value of $E_0$, we predict a vanishing value for $\Omega_{\pm}=0$, and by a further increase of $E_0$ the phonon frequency becomes imaginary $\Omega^2_{\pm}<0$ indicating a structural instability. As a result of this light-induced instability, atomic layers can easily slide to emerge in other stable or metastable staking orders. 

In Fig. \ref{splitting}c, we show the frequency dependence of the normal modes' energies at room temperature showing the non-monotonic profile with a strong dependence on the light frequency. In the sub-gap regime, the phonon frequency drops to zero, then becomes unstable for a range of frequencies around interband transition edge $\hbar\omega=2\mu$. This is because the real part of $\Pi^{ret.}_I$ is enhanced around $\hbar\omega=2\mu$ as depicted in Fig.~\ref{intermode_coupling}c. Further increasing the laser frequency makes $\Omega^2_\pm$ positive and thus stable again. The phonon modes' splitting is stronger at higher laser frequency where $K_I(0)$ becomes less relevant than $K_Z$.
Fig. \ref{splitting}d depicts the temperature dependence of the normal mode frequencies at the interband transition edge $\hbar\omega=2|\mu|$ and for a field amplitude $E_0=0.1$V/nm. The real part of $\Pi^{ret.}_I$ is larger at lower temperatures, making both shear modes unstable. By raising the electronic temperature, phonon modes become stable again, and by a further increase in temperature, the renormalization of phonon frequency starts to converge.
In addition to the field-dependent phonon frequency, we obtain a robust enhancement of phonon linewidth shown in Fig. \ref{splitting}e, due to the photon-mediated amplification of electron-phonon scattering. Finally, we investigate the frequency dependence of critical electric fields $E_\pm$ at which phonon modes $\Omega_\pm$ become unstable. Fig. \ref{splitting}f shows that the critical fields $E_\pm$ increase by raising the laser frequency. 

Considering the nonlinear Raman force, one can further manipulate shear phonon renormalization and its impact on rigid  shear displacement ${\bf Q}_{0} =  \langle {\bf Q}(t)\rangle_{\rm time-average} $ that reads  
\begin{align}
{\bf Q}_{0} = - \frac{1}{\rho}\hat D(\Omega=0) \cdot \mathbfcal{F}^{D}. 
\end{align}
where $\mathbfcal{F}^{D}$ is the displacive Raman shear force in bilayer graphene \cite{rostami_prb_2022}. One can transform to the normal mode basis where the dynamical matrix is diagonal for which one finds the rigid shift  $Q^\pm_0 = {{\cal F}^D_\pm}/{\rho \Omega^2_\pm}$
for two normal shear phonon modes where ${\cal F}^D_\pm$ are the displacive Raman force components along the normal mode vibrational directions. For the case of $\theta=0$, two normal modes $Q_{+}$ and $Q_{-}$ correspond to vibration along $x$ and $y$ direction, respectively. Therefore, the nonlinear Raman force mechanism modulates the light-induced rigid shear displacement via the optically driven renormalization of shear phonon frequency.  

\section{Conclusion and Outlook}
In conclusion, we present a complete quantum theory incorporating coherent dressing of electrons and phonons perturbatively. Unlike Floquet theory, the validity of our approach based on Green's function method is for a wide range of driving field frequencies.
We apply the formal theory to the coherent optical engineering of shear phonons in bilayer graphene. We obtained strong renormalization of shear phonons' frequency that time-resolved spectroscopy of shear phonons can probe in pump-probe experiments \cite{Ulstrup_prl_2014,Cilento_prr_2019,Luo2021,Hein2020,Giorgianni_nc_2022}. In particular, we predict a light-induced non-thermal instability of shear vibration modes that can facilitate nondestructive coherent engineering lattice structure in layered materials. Our theory can be applied to other types of phonon modes in heterostructures of layered materials, which involve relative twists of layers. Having a coherent control of shear phonon dynamics provides an optical switching of polar metals, moiré ferroelectrics, and superconductivity in the heterostructures of layered quantum materials \cite{Zheng_nature_2020,Woods_nc_2021,Yasuda_science_2021,Vizner_science_2021,Fei_nature_2018, Sharma_sciad_2019, Zhou_jjap_2020,Qin_prl_2021,Liang_prx_2022}. For intense incident laser, there is a saturation effect of the light-induced displacement usually observed in experimental measurements of coherent phonon displacement amplitude. This effect is due to the saturation of the optical absorption that can be explained via a saturable absorption process described by the third-order optical conductivity and a nonlinear force forth-order in the electric field amplitude, e.g., $\mathbfcal{F}\propto EE^\ast E E^\ast$. The saturation effect analysis is beyond the scope of this manuscript and will be discussed elsewhere.

\section*{Acknowledgment}
I acknowledge the support from the Swedish Research Council (VR Starting Grant No. 2018-04252). Nordita is partially supported by Nordforsk. I am grateful to E. Cappelluti and J. Hofmann for the valuable discussion and feedback.

\bibliography{refs.bib} 
\onecolumngrid
\pagebreak
\appendix

\section{Instantaneous susceptibility} \label{app:inst} 
Considering the contribution of $\Delta^{(2)}_{abcd}$, the instantaneous coupling consists of three contributions
\begin{align}
\bar\rchi^{ins.}_{abcd} (\omega_1,\omega_2) = \bar\rchi^{\rm triangle}_{abcd} (\omega_1,\omega_2)+\bar\rchi^{\rm bubble-\gamma}_{abcd} (\omega_1,\omega_2)+\bar\rchi^{\rm bubble-\Theta}_{abcd} (\omega_1,\omega_2)
\end{align}
In the following subsections, we calculate the values of each diagram for the instantaneous susceptibility. 
\subsection{Calculation of $\bar\rchi^{\rm triangle}_{abcd}$ for the diagram depicted in Fig. \ref{fig:diagrams}a } 
The triangle diagram Fig.~\ref{fig:diagrams}a can be written in terms of electronic Green's function $\hat G({\bf k},ik_n)$ and two-phonon-electron matrix-element $\hat {\cal M}^{(2)}_{ab}$ and paramagnetic current operator $\hat j_c, \hat j_d$: 
\begin{align}
\rchi_{abcd}(i\omega_{m_1},i\omega_{m_2}) = \sum_{\cal P} \frac{1}{S}\sum_{\bf k}
\frac{1}{\beta}\sum_{ik_n}
{\rm Tr}\left[\hat {\cal M}^{(2)}_{ab}({\bf k}) \hat G({\bf k},ik_n) \hat j_c({\bf k}) \hat G({\bf k},ik_n+i\omega_{m_1})\hat j_d({\bf k}) \hat G({\bf k},ik_n+i\omega_{m_1}+i\omega_{m_2})\right]
\end{align}
where the trace operator ${\rm Tr}[\dots]$ sum over all spinor degree of freedom, $\beta =1/k_{\rm B}T_e$, $ik_n$ ($i\omega_m$) stands for the fermionic (bosonic) Matsubara frequency.
The intrinsic permutation symmetry is enforced by $\sum_{\cal P}$ for the exchange of photon frequencies and corresponding tensorial index: $(c,m_1) \leftrightarrow (d,m_2)$. 
From now on, we adopt a short-hand notation $ik_n\to n$ and $i\omega_m\to m$ for the sake of simplicity.  
The electronic Green's function is given as follows 
\begin{align}
\hat G({\bf k},ik_n) = [ik_n - \hat H_{\bf k}]^{-1}. 
\end{align}
Because of the inversion symmetry, the response tensor elements with odd Cartesian index $x$ and $y$ vanishes
$\rchi_{xxxy}=\rchi_{xxyx}, =\rchi_{yyxy} = \rchi_{yyyx}=\rchi_{xyxx}=\rchi_{xyyy}=\rchi_{yxxx}=\chi_{yxyy}=0$. This symmetry consideration is confirmed by an explicit calculation based on the low-energy two-band model. 
The remaining tensor elements are also related to each other due to the rotation symmetry of the system: 
\begin{align}
-\rchi_{xxxx} = -\rchi_{yyyy}=
 \rchi_{xxyy} = \rchi_{yyxx} =
 \rchi_{xyyx} = \rchi_{xyxy} =
 \rchi_{yxyx} = \rchi_{yxxy} = \rchi_1.
\end{align}
After performing the integration on the azimuthal angle of electronic wave vector $\bf k$ and using the low-energy dispersion $\epsilon_{\bf k} = \hbar^2 k^2/2m$ and $kdk= (m/\hbar^2)d\epsilon$ we find 
\begin{align}
\chi_1(m_1,m_2)&= \left(\frac{N_f {\cal M}^{(2)}}{2\pi}\right)\left (\frac{e}{m}\right)^2 \left(\frac{m}{\hbar^2}\right) \int^\infty_0 d \epsilon \frac{1}{\beta}\sum_n
\nonumber\\ &
\frac{8 \epsilon^2 \xi(n) (2\epsilon^2-\xi(m_1+n)^2-\xi(m_2+n)^2) \xi(m_1+m_2+n)}{\left(\epsilon^2-\xi(n)^2\right) \left(\epsilon^2-\xi(m_1+n)^2\right) \left(\epsilon^2-\xi(m_2+n)^2\right)
  \left(\epsilon^2-\xi(m_1+m_2+n)^2\right)}.
\end{align}
where $\xi(n)= \mu+n$. After performing Matsubara summation, integrating over $\epsilon$ at zero temperature and analytical continuation $m_i\to \omega_i+i0^+$, we find 
\begin{align}
\rchi_1(\omega_1,\omega_2) = \frac{N_f {\cal M}^{(2)} e^2}{4\pi \hbar^2} \bigg\{ A_1 \ln[4\epsilon^2-\omega^2_1] + A_2 \ln[4\epsilon^2-\omega^2_2] + A_3 \ln[4\epsilon^2-(\omega_1+\omega_2)^2] \bigg\}^{\epsilon \to\infty}_{\epsilon\to\mu}.
\end{align}
Here by $\omega_i$ we mean $\hbar\omega_i+i0^+$ and $A_i$ factors read  
\begin{align}
A_1 = \frac{\omega_1(\omega_1+2\omega_2)}{\omega_2(\omega_1+\omega_2)}~,~~~
A_2 = \frac{\omega_2(\omega_2+2\omega_1)}{\omega_1(\omega_1+\omega_2)}~,~~~
A_3 =-1- (A_1+A_2).
\end{align} 
By subtracting the zero-frequency contribution and after some simplifications, we find 
\begin{align}
\rchi_1(\omega_1,\omega_2)-\rchi_1(0,0) = \frac{N_f {\cal M}^{(2)} e^2}{4\pi \hbar^2} \bigg\{ A_1 \ln\left[\frac{4\epsilon^2-\omega^2_1}{4\epsilon^2-(\omega_1+\omega_2)^2}\right] + A_2 \ln\left[\frac{4\epsilon^2-\omega^2_2}{4\epsilon^2-(\omega_1+\omega_2)^2}\right] - \ln \left[\frac{4\epsilon^2-(\omega_1+\omega_2)^2}{4\epsilon^2}\right] \bigg\}^{\epsilon \to\infty}_{\epsilon\to\mu}.
\end{align}
Eventually, we obtain $\bar\chi^{\rm triangle}_{1}(\omega_1,\omega_2) =\chi_1(\omega_1,\omega_2)-\chi_1(0,0)$ as follows
\begin{align}
\bar\rchi^{\rm triangle}_{1}(\omega_1,\omega_2)= \frac{ N_f {\cal M}^{(2)} e^2}{4\pi \hbar^2} \bigg\{\ln \left[1-\frac{(\omega_1+\omega_2)^2}{4\mu^2}\right] -A_1 \ln\left[\frac{4\mu^2-\omega^2_1}{4\mu^2-(\omega_1+\omega_2)^2}\right] 
 - A_2 \ln\left[\frac{4\mu^2-\omega^2_2}{4\mu^2-(\omega_1+\omega_2)^2}\right] \bigg\}.
\end{align}
\subsection{Calculation of $\bar\rchi^{\rm bubble-\gamma}_{abcd}$  for the diagram depicted in Fig. \ref{fig:diagrams}b}
The bubble diagram Fig.~\ref{fig:diagrams}b can be written in terms of electronic Green's function $\hat G({\bf k},n)$, electron-phonon matrix-element $\hat {\cal M}^{(2)}_{ab}$ and the Raman vertex $\hat \gamma_{cd}$: 
\begin{align}
\rchi_{abcd}(m_1,m_2) = - \frac{1}{S}\sum_{\bf k}
\frac{1}{\beta}\sum_{n}
{\rm Tr}\left[\hat {\cal M}^{(2)}_{ab}({\bf k}) \hat G({\bf k},n) \hat \gamma_{cd}({\bf k}) \hat G({\bf k},n+m_1+m_2)\right]
\end{align}
The overall minus sign originates from the standard rules of Feynman diagrams \cite{Mahan}, also see \cite{Rostami_ap_2021}. Similar to the previous diagram, we have $\rchi_{xxxy}=\rchi_{xxyx} =\rchi_{yyxy} = \rchi_{yyyx}=\rchi_{xyxx}=\rchi_{xyyy}=\rchi_{yxxx}=\rchi_{yxyy}=0$. The other non-vanishing tensor elements read
\begin{align}
-\rchi_{xxxx} = -\rchi_{yyyy}=
 \rchi_{xxyy} = \rchi_{yyxx} =
 \rchi_{xyyx} = \rchi_{xyxy} =
 \rchi_{yxyx} = \rchi_{yxxy} = \rchi_2.
\end{align}
After performing the integration on the azimuthal angle of electronic wave vector $\bf k$ and using the low-energy dispersion $\epsilon_{\bf k} = \hbar^2 k^2/2m$ and $kdk= (m/\hbar^2)d\epsilon$ we find 
\begin{align}
\chi_2(m_1,m_2)= N_f\frac{m}{\hbar^2}\frac{e^2 {\cal M}^{(2)}}{2\pi m} \int^\infty_0 d\epsilon \frac{1}{\beta}\sum_n \frac{2 \xi(n) \xi(m_1+m_2+n)}{\left(\epsilon^2-\xi(n)^2\right) \left(\epsilon^2-\xi(m_1+m_2+n)^2\right)}.
\end{align}
After performing the summation on the Matsubara frequency $n$ and subtracting the zero-frequency contribution, we find  
\begin{align}
\chi_2(\omega_1,\omega_2)- \chi_2(0,0)
= -\frac{ N_f {\cal M}^{(2)} e^2}{8\pi\hbar^2} \Big\{\ln[\frac{4\epsilon^2-(\omega_1+\omega_2)^2}{4\epsilon^2}]\Big\}^{\epsilon\to\infty }_{\epsilon\to\mu}.
\end{align}
Finally, we obtain 
\begin{align}
\bar\rchi^{\rm bubble-\gamma}_{2}(\omega_1,\omega_2)
=  \frac{N_f {\cal M}^{(2)} e^2}{8\pi\hbar^2} \ln\left [1-\frac{(\omega_1+\omega_2)^2}{4\mu^2} \right ].
\end{align}
\subsection{Calculation of $\bar\rchi^{\rm bubble-\Theta}_{abcd}$  for the diagram depicted in Fig. \ref{fig:diagrams}c}
The bubble diagram Fig.~\ref{fig:diagrams}c can be written in terms of electronic Green's function $\hat G({\bf k},n)$, 1photon-electron-phonon vertex $\hat \Theta^{(2)}_{abc}$ and the paramagnetic current $\hat j_{d}$. Considering the permutation symmetry, we have  
\begin{align}
\chi_{abcd}(m_1,m_2) &= - \frac{1}{2S}\sum_{\bf k}
\frac{1}{\beta}\sum_{n}
{\rm Tr}\left[\hat \Theta^{(2)}_{abc}({\bf k}) \hat G({\bf k},n) \hat j_d({\bf k}) \hat G({\bf k},n+m_2)\right]
\nonumber\\
&- \frac{1}{2S}\sum_{\bf k}
\frac{1}{\beta}\sum_{n}
{\rm Tr}\left[\hat \Theta^{(2)}_{abd}({\bf k})  \hat G({\bf k},n) \hat j_c({\bf k}) \hat G({\bf k},n+m_1)\right].
\end{align}
Using the isotropic approximation for the PEP vertex given in Eq. (6) and after performing the integration on the azimuthal angle of electronic wave vector $\bf k$, we obtain a vanishing result for all tensor elements. Therefore, within our low-energy model analysis, the mix of photon-electron-phonon coupling does not contribute to the Raman force:
\begin{align}
\bar\rchi^{\rm bubble-\Theta}_{abcd} (\omega_1,\omega_2)=0. 
\end{align}
\subsection{The sum of all diagrams for the instantaneous coupling}
Similar to the Raman force case, we obtain $\bar\rchi^{ins.}(\omega_1,\omega_2)=\bar\rchi_1(\omega_1,\omega_2)+\bar\rchi_2(\omega_1,\omega_2)$.  One main difference is that instead of ${\cal M}^{(1)}$ we have ${\cal M}^{(2)}$: 
\begin{align}
\bar\rchi^{ins.}_{xxxx}(\omega_1,\omega_2)&= \frac{ N_f {\cal M}^{(2)} e^2}{4\pi \hbar^2} \bigg\{ \frac{3}{2}\ln \left[1-\frac{(\omega_1+\omega_2)^2}{4\mu^2}\right]
-\frac{\omega_1(\omega_1+2\omega_2)}{\omega_2(\omega_1+\omega_2)} \ln\left[\frac{4\mu^2-\omega^2_1}{4\mu^2-(\omega_1+\omega_2)^2}\right] 
\nonumber\\& - \frac{\omega_2(\omega_2+2\omega_1)}{\omega_1(\omega_1+\omega_2)} \ln\left[\frac{4\mu^2-\omega^2_2}{4\mu^2-(\omega_1+\omega_2)^2}\right] \bigg\}.
\end{align}
Finally, by considering a linear polarized incident electric field ${\bf E}(t) = E_0 (\hat{\bf x}\cos\theta+\hat{\bf y}\sin\theta) e^{-i\omega t}+c.c$, we find 
\begin{align}
    \hat {\cal G}^{ins.}(\omega_1,\omega_2) =  \Pi^{ins.}_{xxxx}(\omega_1,\omega_2)    E^2_0 
    \begin{bmatrix}
    -\cos(2\theta) &\sin(2\theta) \\[5pt] \sin(2\theta) & \cos(2\theta) 
    \end{bmatrix}.
\end{align}
The rectified part of $ \hat {\cal G}^{ins.}$ is obtain after setting $\omega_1=\hbar(\omega+i0^+)$ and $\omega_2=\hbar(-\omega+i0^+)$ where $\omega$ is the incident laser frequency and $\Gamma_e$ stands for the effective scattering rate of  electrons.  
\section{Retarded susceptibility}\label{app:ret} 
The retarded coupling is given in terms of five different diagram 
\begin{align}
\bar\rchi^{ret.}_{abcd} (\omega_1,\omega_2,\omega_3) 
&= \bar\rchi^{\rm square}_{abcd} (\omega_1,\omega_2,\omega_3)+\bar\rchi^{\rm triangle-\gamma}_{abcd} (\omega_1,\omega_2,\omega_3)
+\bar\rchi^{\rm triangle-\Theta}_{abcd} (\omega_1,\omega_2,\omega_3)
\nonumber\\
&+\bar\rchi^{\rm bubble-\Theta}_{abcd} (\omega_1,\omega_2,\omega_3)
+\bar\rchi^{\rm bubble-\Delta}_{abcd} (\omega_1,\omega_2,\omega_3). 
\end{align}
In the following, we calculate the explicit expression of each contribution using standard Kubo's formalism at zero electronic temperature.  
\subsection{Calculation of $\bar\rchi^{\rm square}_{abcd}$  for the diagram depicted in Fig. \ref{fig:diagrams}e}
The square diagram Fig.~\ref{fig:diagrams}e can be written in terms of electronic Green's function $\hat G({\bf k},ik_n)$ and two-phonon-electron matrix-element $\hat {\cal M}^{(1)}_{a}$, $\hat {\cal M}^{(1)}_{b}$ and paramagnetic current operators $\hat j_c, \hat j_d$: 
\begin{align}
\rchi_{abcd}(m_1,m_2,m_3) &= \frac{1}{3!}\sum_{\cal P} \frac{1}{S}\sum_{\bf k}
\frac{1}{\beta}\sum_{n}
{\rm Tr}\Big[
\hat {\cal M}^{(1)}_{a} ({\bf k})
\hat G({\bf k},n)
\hat {\cal M}^{(1)}_{b} ({\bf k})
\hat G({\bf k},n+m_3) 
\hat j_c({\bf k}) 
\hat G({\bf k},n+m_3+m_1)
\nonumber\\& 
\hat j_d({\bf k}) 
\hat G({\bf k},n+m_3+m_1+m_2) 
\Big].
\end{align}
Note that $\sum_{\cal P}$ stands to ensure the intrinsic permutation symmetry. Because of the inversion symmetry, the response tensor elements with odd Cartesian index $x$ and $y$ vanishes
$\rchi_{xxxy} =\rchi_{xxyx} =
\rchi_{yyxy} = \rchi_{yyyx}=
\rchi_{xyxx}=\rchi_{xyyy}=
\rchi_{yxxx}=\rchi_{yxyy}=0$. 
Accordingly, there are only four independent tensor elements 
\begin{align}
\rchi_{xxxx} = \rchi_{yyyy},~
\rchi_{xxyy} = \rchi_{yyxx},~
\rchi_{xyxy} = \rchi_{yxyx},~
\rchi_{xyyx} = \rchi_{yxxy}.
\end{align}
By performing a straightforward algebra similar to what was discussed in the previous subsection, one can obtain the four non-vanishing tensor elements in the following form: 
\begin{align}
 &\rchi_{xxxx}(\omega_1,\omega_2,\omega_3) =\rchi_{xxyy}(\omega_1,\omega_2,\omega_3) 
 = \frac{N_f}{24\pi} \left(\frac{e{\cal M}^{(1)}}{\hbar}  \right)^2 \times\nonumber\\&
 \Bigg\{
   -\frac{4 \omega _1 \omega _2 \left(\omega _1^2+\left(\omega _2+\omega _3\right) \omega _1+\omega _3 \left(\omega
   _2+\omega _3\right)\right)}{\left(\omega _1+\omega _2\right) \omega _3 \left(\omega
   _1+\omega _3\right) \left(\omega _2+\omega _3\right) \left(\omega _1+\omega _2+\omega _3\right)} \ln \left[\frac{2|\mu|-\omega _1}{2|\mu|+\omega _1}\right]
   \nonumber\\&
    -\frac{4 \omega _1 \omega _2 \left(\omega _2^2+\omega _3 \omega _2+\omega _3^2+\omega _1 \left(\omega _2+\omega _3\right)\right)}{\left(\omega _1+\omega _2\right) \omega _3 \left(\omega _1+\omega _3\right) \left(\omega _2+\omega _3\right)
   \left(\omega _1+\omega _2+\omega _3\right)}
   \ln\left[\frac{2|\mu|-\omega_2}{2|\mu|+\omega_2}\right]
   \nonumber\\&
   +\frac{4 \omega _3 \left(\left(\omega _2+\omega _3\right)
   \omega _1^2+\left(\omega _2^2+3 \omega _3 \omega _2+\omega _3^2\right) \omega _1+\omega _2 \omega _3 \left(\omega _2+\omega _3\right)\right)}{\omega _1 \omega _2 \left(\omega _1+\omega _2\right) \left(\omega _1+\omega _3\right)
   \left(\omega _2+\omega _3\right)}
   \ln \left[ \frac{2|\mu|-\omega _3}{2|\mu|+\omega _3} \right]
   \nonumber\\&
   -\frac{4 \left(\omega _1+\omega _3\right) \left(\omega _1 \left(\omega _3-\omega _2\right)+\omega _3 \left(\omega
   _2+\omega _3\right)\right)}{\omega _1 \omega _2 \omega _3
   \left(\omega _1+\omega _2+\omega _3\right)}
   \ln\left[ \frac{2|\mu|-\omega _1-\omega _3}{2|\mu|+\omega _1+\omega _3}\right]
   \nonumber\\&
   -\frac{4 \left(\omega _2+\omega _3\right) \left(\omega _1 \left(\omega _3-\omega _2\right)+\omega _3 \left(\omega
   _2+\omega _3\right)\right)}{\omega _1 \omega _2 \omega _3
   \left(\omega _1+\omega _2+\omega _3\right)} 
   \ln\left[ \frac{2|\mu|-\omega _2-\omega _3}{2|\mu|+\omega _2+\omega _3}\right]
   \nonumber\\&
   +\frac{4 \omega _3 \left(\omega _1^3+2 \left(\omega _2+\omega _3\right) \omega _1^2+\left(2 \omega _2^2+3 \omega _3 \omega
   _2+\omega _3^2\right) \omega _1+\omega _2 \left(\omega _2+\omega _3\right){}^2\right)}{\omega _1 \omega _2 \left(\omega _1+\omega _2\right) \left(\omega _1+\omega _3\right) \left(\omega
   _2+\omega _3\right)} 
   \ln\left[\frac{2|\mu|-\omega _1-\omega _2-\omega _3}{2|\mu|+\omega _1+\omega _2+\omega _3}\right]
   \Bigg\}, 
\end{align}
and 
\begin{align}
&\rchi_{xyxy} (\omega_1,\omega_2,\omega_3)= - \rchi_{xyyx}(\omega_1,\omega_2,\omega_3)= 
\frac{N_f}{24\pi} \left(\frac{e{\cal M}^{(1)}}{\hbar}  \right)^2 
\Bigg\{ 
   \frac{4 \omega _1 \left(\omega _1+\omega _2+2 \omega
   _3\right) }{\omega _2 \left(\omega _1+\omega _2+\omega _3\right) \omega _3}
      \ln\left[ \frac{2|\mu|-\omega _1}{2|\mu|+\omega _1}\right]
   %
   \nonumber\\&
   -\frac{4 \omega _2 \left(\omega _1+\omega _2+2
   \omega _3\right) }{\omega _1 \left(\omega _1+\omega _2+\omega _3\right) \omega
   _3}
      \ln\left[ \frac{2|\mu|-\omega _2}{2|\mu|+\omega _2}\right]
   %
   %
+\frac{4 \left(\omega _1-\omega _2\right) \omega _3 }{\omega _1 \omega _2 \left(\omega
   _1+\omega _2+\omega _3\right)}
    \ln\left[ \frac{2|\mu|-\omega _3}{2|\mu|+\omega _3}\right]
   %
   \nonumber\\&
   -\frac{4 \left(\omega _1^2-\omega _2^2\right) \left(\omega _1+\omega _2+2 \omega _3\right) }{\omega _1 \omega _2 \left(\omega _1+\omega _2+\omega _3\right) \omega _3}
      \ln\left[ \frac{2|\mu|-\omega _1-\omega _2}{2|\mu|+\omega _1+\omega _2}\right]
   %
   \nonumber\\&
   -\frac{4
   \left(\omega _1+\omega _3\right) \left(\omega _1^2+\left(\omega _2+\omega _3\right) \omega _1-\omega _2 \omega _3\right) }{\omega _1 \omega _2 \left(\omega _1+\omega _2+\omega _3\right) \omega _3}
      \ln\left[ \frac{2|\mu|-\omega _1-\omega _3}{2|\mu|+\omega _1+\omega _3}\right]
   %
   \nonumber\\&
   +\frac{4 \left(\omega
   _2+\omega _3\right) \left(\omega _1 \left(\omega _2-\omega _3\right)+\omega _2 \left(\omega _2+\omega _3\right)\right) }{\omega _1 \omega _2 \left(\omega _1+\omega _2+\omega _3\right) \omega _3}
         \ln\left[ \frac{2|\mu|-\omega _2-\omega _3}{2|\mu|+\omega _2+\omega _3}\right]
      \nonumber\\&
   +\frac{4 \left(\omega _1-\omega _2\right) \left(\omega _1+\omega _2+\omega _3\right) }{\omega _1 \omega _2 \omega _3}
      \ln\left[\frac{2|\mu|-\omega _1-\omega _2-\omega _3}{2|\mu|+\omega _1+\omega _2+\omega _3}\right]
   %
 \Bigg\}. 
\end{align}
For the short hand notation we adapt $\omega_i$ for $\hbar(\omega_i +i0^+)$ in the above relations.
\subsection{Calculation of $\bar\rchi^{\rm triangle-\gamma}_{abcd}$  for the diagram depicted in Fig. \ref{fig:diagrams}f}
The triangle diagram Fig.~\ref{fig:diagrams}f can be written in terms of electronic Green's function $\hat G({\bf k},ik_n)$ and electron-phonon matrix-element $\hat {\cal M}^{(1)}_{a}$, $\hat {\cal M}^{(1)}_{b}$ and diamagnetic current operator $\hat \gamma_{cd}$: 
\begin{align}
\rchi_{abcd}(m_1,m_2,m_3) = -\sum_{\cal P} \frac{1}{S}\sum_{\bf k}
\frac{1}{\beta}\sum_{n}
{\rm Tr}\left[
\hat {\cal M}^{(1)}_{a} ({\bf k})
\hat G({\bf k},n)
\hat {\cal M}^{(1)}_{b} 
\hat G({\bf k},n+m_1) 
\hat \gamma_{cd}({\bf k}) 
\hat G({\bf k},n+m_1+m_2+m_3)
\right] 
\end{align}
Using the isotropic approximation model Hamiltonian and after performing the integration on the azimuthal angle of electronic wave vector $\bf k$, we obtain a vanishing result for all tensor elements. Therefore, within our low-energy model analysis, we have 
\begin{align}
\rchi^{\rm triangle-\gamma}_{abcd}(\omega_1,\omega_2,\omega_3)=0.   
\end{align}
\subsection{Calculation of $\bar\rchi^{\rm triangle-\Theta}_{abcd}$  for the diagram depicted in Fig. \ref{fig:diagrams}g}
The triangle diagram Fig.~\ref{fig:diagrams}g can be written in terms of electronic Green's function $\hat G({\bf k},ik_n)$ and photon-electron-phonon matrix-element $\hat {\Theta}^{(1)}_{ac}$, electron-phonon matrix-element $\hat {\cal M}^{(1)}_{b}$ and paramagnetic current operator $\hat j_{d}$: 
\begin{align}
\rchi_{abcd}(m_1,m_2,m_3) 
&=  
\sum_{\cal P}\frac{1}{2S}\sum_{\bf k}
\frac{1}{\beta}\sum_{n}
{\rm Tr}\left[
\hat \Theta^{(1)}_{ac}({\bf k})  
\hat G({\bf k},n)
\hat {\cal M}^{(1)}_{b}({\bf k})  
\hat G({\bf k},n+m_1) 
\hat j_{d}({\bf k}) 
\hat G({\bf k},n+m_1+m_3)
\right] 
\nonumber\\ 
&+ 
\sum_{\cal P}\frac{1}{2S}\sum_{\bf k}
\frac{1}{\beta}\sum_{n}
{\rm Tr}\left[
\hat \Theta^{(1)}_{ad}({\bf k})  
\hat G({\bf k},n)
\hat {\cal M}^{(1)}_{b}({\bf k})  
\hat G({\bf k},n+m_1) 
\hat j_{c}({\bf k}) 
\hat G({\bf k},n+m_1+m_2)
\right]. 
\end{align}
Using the isotropic approximation model Hamiltonian and after performing the integration on the azimuthal angle of electronic wave vector $\bf k$, we obtain a vanishing result for all tensor elements. Therefore, within our low-energy model analysis, we have 
\begin{align}
\rchi^{\rm triangle-\Theta}_{abcd}(\omega_1,\omega_2,\omega_3)=0.   
\end{align}
\subsection{Calculation of $\bar\rchi^{\rm bubble-\Theta}_{abcd}$  for the diagram depicted in Fig. \ref{fig:diagrams}h}
The triangle diagram Fig.~\ref{fig:diagrams}h can be written in terms of electronic Green's function $\hat G({\bf k},ik_n)$ and photon-electron-phonon matrix-element $\hat { \Theta}^{(1)}_{ac}$: 
\begin{align}
\rchi_{abcd}(m_1,m_2,m_3) &= \sum_{\cal P} \frac{1}{2S}\sum_{\bf k}
\frac{1}{\beta}\sum_{n}
{\rm Tr}\left[
\hat \Theta^{(1)}_{ac} ({\bf k})
\hat G({\bf k},n)
\hat \Theta^{(1)}_{bd} ({\bf k})
\hat G({\bf k},n+m_2+m_3) 
\right] 
\nonumber\\
&+ \sum_{\cal P}\frac{1}{2S}\sum_{\bf k}
\frac{1}{\beta}\sum_{n}
{\rm Tr}\left[
\hat \Theta^{(1)}_{ad} ({\bf k})
\hat G({\bf k},n)
\hat \Theta^{(1)}_{bc} ({\bf k})
\hat G({\bf k},n+m_1+m_3) 
\right].
\end{align}
Similar to the square diagram the only non-vanishing tensor elements are given by 
$\rchi_{xxxx} = \rchi_{yyyy},~
\rchi_{xxyy} = \rchi_{yyxx},~
\rchi_{xyxy} = \rchi_{yxyx},~
\rchi_{xyyx} = \rchi_{yxxy}.$ The straightforward algebra similar to what was discussed earlier, one can obtain the four non-vanishing tensor elements in the following form: 
\begin{align}
&\bar\rchi_{xxxx} (\omega_1,\omega_2,\omega_3) = C_\Theta 
\Bigg\{ 
\ln
\left [ 
1-\frac{(\omega_1+\omega_2)^2}{4\mu^2}
\right] 
+
\ln
\left [ 
1-\frac{(\omega_1+\omega_3)^2}{4\mu^2}
\right] 
+
\ln
\left [ 
1-\frac{(\omega_2+\omega_3)^2}{4\mu^2}
\right] 
\Bigg\}, 
\\
&\bar\rchi_{xxyy} (\omega_1,\omega_2,\omega_3) = C_\Theta 
\Bigg\{ 
-\ln
\left [ 
1-\frac{(\omega_1+\omega_2)^2}{4\mu^2}
\right] 
+
\ln
\left [ 
1-\frac{(\omega_1+\omega_3)^2}{4\mu^2}
\right] 
+
\ln
\left [ 
1-\frac{(\omega_2+\omega_3)^2}{4\mu^2}
\right] 
\Bigg\},
\\
&\bar\rchi_{xyyx}  (\omega_1,\omega_2,\omega_3)= C_\Theta 
\Bigg\{ 
\ln
\left [ 
1-\frac{(\omega_1+\omega_2)^2}{4\mu^2}
\right] 
-
\ln
\left [ 
1-\frac{(\omega_1+\omega_3)^2}{4\mu^2}
\right] 
+
\ln
\left [ 
1-\frac{(\omega_2+\omega_3)^2}{4\mu^2}
\right] 
\Bigg\}, 
\\
&\bar\rchi_{xyxy}  (\omega_1,\omega_2,\omega_3)=C_\Theta 
\Bigg\{ 
\ln
\left [ 
1-\frac{(\omega_1+\omega_2)^2}{4\mu^2}
\right] 
+
\ln
\left [ 
1-\frac{(\omega_1+\omega_3)^2}{4\mu^2}
\right] 
-
\ln
\left [ 
1-\frac{(\omega_2+\omega_3)^2}{4\mu^2}
\right] 
\Bigg\}. 
\end{align}
where 
\begin{align}
  C_\Theta = \frac{N_f m [\Theta^{(1)}]^2}{24\pi \hbar^2} .   
\end{align}
\subsection{Calculation of $\bar\rchi^{\rm bubble-\Delta}_{abcd}$  for the diagram depicted in Fig. \ref{fig:diagrams}i}
The triangle diagram Fig.~\ref{fig:diagrams}i can be written in terms of electronic Green's function $\hat G({\bf k},ik_n)$ and photon-electron-phonon matrix-element $\hat \Delta^{(1)}_{acd}$ and electron-phonon matrix-element $\hat {\cal M}^{(1)}_{b}$: 
\begin{align}
\chi_{5,abcd}(m_1,m_2,m_3) &= \sum_{\cal P} \frac{1}{2S}\sum_{\bf k}
\frac{1}{\beta}\sum_{n}
{\rm Tr}\left[
\hat \Delta^{(1)}_{acd} ({\bf k})
\hat G({\bf k},n)
\hat {\cal M}^{(1)}_{b}({\bf k})
\hat G({\bf k},n+m_3) 
\right] 
\end{align}
Similar to the square diagram the only non-vanishing tensor elements are given by 
$\rchi_{xxxx} = \rchi_{yyyy},~
\rchi_{xxyy} = \rchi_{yyxx},~
\rchi_{xyxy} = \rchi_{yxyx},~
\rchi_{xyyx} = \rchi_{yxxy}.$ The straightforward algebra similar to what was discussed earlier, one can obtain the four non-vanishing tensor elements in the following form: 
\begin{align}
&\bar\rchi_{xxxx}(\omega_1,\omega_2,\omega_3) = -\frac{3 C_\Delta}{2} 
\Bigg\{ 
\ln
\left [ 
1-\frac{\omega_1^2}{4\mu^2}
\right] 
+
\ln
\left [ 
1-\frac{\omega_2^2}{4\mu^2}
\right]  
+
\ln
\left [ 
1-\frac{\omega_3^2}{4\mu^2}
\right] 
\Bigg\},
\\
&\bar\rchi_{xxyy}(\omega_1,\omega_2,\omega_3) =-\frac{C_\Delta}{2} 
\Bigg\{ 
-\ln
\left [ 
1-\frac{\omega_1^2}{4\mu^2}
\right] 
-
\ln
\left [ 
1-\frac{\omega_2^2}{4\mu^2}
\right]  
+
\ln
\left [ 
1-\frac{\omega_3^2}{4\mu^2}
\right] 
\Bigg\},
\\
&\bar\rchi_{xyxy}(\omega_1,\omega_2,\omega_3) = -\frac{C_\Delta}{2} 
\Bigg\{ 
\ln
\left [ 
1-\frac{\omega_1^2}{4\mu^2}
\right] 
-
\ln
\left [ 
1-\frac{\omega_2^2}{4\mu^2}
\right]  
-
\ln
\left [ 
1-\frac{\omega_3^2}{4\mu^2}
\right] 
\Bigg\}, 
\\
&\bar\rchi_{xyyx}(\omega_1,\omega_2,\omega_3) =-\frac{C_\Delta}{2} 
\Bigg\{ 
-\ln
\left [ 
1-\frac{\omega_1^2}{4\mu^2}
\right] 
+
\ln
\left [ 
1-\frac{\omega_2^2}{4\mu^2}
\right]  
-
\ln
\left [ 
1-\frac{\omega_3^2}{4\mu^2}
\right] 
\Bigg\}. 
\end{align}
where 
\begin{align}
  C_\Delta =   \frac{ N_f m \Delta^{(1)} {\cal M}^{(1)}}{24\pi\hbar^2}.   
\end{align}
Since $\Delta^{(1)} {\cal M}^{(1)}= [\Theta^{(1)}]^2$, we have $  C_\Delta=  C_\Theta$. 
\end{document}